\PassOptionsToPackage{sort&compress}{natbib}
\documentclass[preprint,11pt]{elsarticle}

\usepackage{geometry}
\journal{Journal of \LaTeX\ Templates}

\usepackage{multirow}
\usepackage{multicol,lipsum}
\usepackage{color, colortbl}
\usepackage{dblfloatfix}

\usepackage{threeparttable}
\usepackage{tikz}

\newcommand{\ra}[1]{\renewcommand{\arraystretch}{#1}}

\usepackage{amsmath,amsthm,amssymb,amsthm,amsfonts,mathtools}

\usepackage{graphicx}
\usepackage{eqnarray}
\usepackage{url}
\usepackage{booktabs,multirow,array}
\usepackage{ragged2e}
\usepackage{enumerate}
\graphicspath{{Figs2/}}
\newcommand{\mi}{i} 
\usepackage{bm}

\makeatletter
\let\save@mathaccent\mathaccent
\newcommand*\if@single[3]{%
	\setbox0\hbox{${\mathaccent"0362{#1}}^H$}%
	\setbox2\hbox{${\mathaccent"0362{\kern0pt#1}}^H$}%
	\ifdim\ht0=\ht2 #3\else #2\fi
}
\newcommand*\rel@kern[1]{\kern#1\dimexpr\macc@kerna}
\newcommand*\widebar[1]{\@ifnextchar^{{\wide@bar{#1}{0}}}{\wide@bar{#1}{1}}}
\newcommand*\wide@bar[2]{\if@single{#1}{\wide@bar@{#1}{#2}{1}}{\wide@bar@{#1}{#2}{2}}}
\newcommand*\wide@bar@[3]{%
	\begingroup
	\def\mathaccent##1##2{%
		\let\mathaccent\save@mathaccent
		\if#32 \let\macc@nucleus\first@char \fi
		\setbox\z@\hbox{$\macc@style{\macc@nucleus}_{}$}%
		\setbox\tw@\hbox{$\macc@style{\macc@nucleus}{}_{}$}%
		\dimen@\wd\tw@
		\advance\dimen@-\wd\z@
		\divide\dimen@ 3
		\@tempdima\wd\tw@
		\advance\@tempdima-\scriptspace
		\divide\@tempdima 10
		\advance\dimen@-\@tempdima
		\ifdim\dimen@>\z@ \dimen@0pt\fi
		\rel@kern{0.6}\kern-\dimen@
		\if#31
		\overline{\rel@kern{-0.6}\kern\dimen@\macc@nucleus\rel@kern{0.4}\kern\dimen@}%
		\advance\dimen@0.4\dimexpr\macc@kerna
		\let\final@kern#2%
		\ifdim\dimen@<\z@ \let\final@kern1\fi
		\if\final@kern1 \kern-\dimen@\fi
		\else
		\overline{\rel@kern{-0.6}\kern\dimen@#1}%
		\fi
	}%
	\macc@depth\@ne
	\let\math@bgroup\@empty \let\math@egroup\macc@set@skewchar
	\mathsurround\z@ \frozen@everymath{\mathgroup\macc@group\relax}%
	\macc@set@skewchar\relax
	\let\mathaccentV\macc@nested@a
	\if#31
	\macc@nested@a\relax111{#1}%
	\else
	\def\gobble@till@marker##1\endmarker{}%
	\futurelet\first@char\gobble@till@marker#1\endmarker
	\ifcat\noexpand\first@char A\else
	\def\first@char{}%
	\fi
	\macc@nested@a\relax111{\first@char}%
	\fi
	\endgroup
}
\makeatother
\usepackage{accents}
\usepackage{upgreek}
\usepackage{ragged2e}
\usepackage{stmaryrd}

\theoremstyle{plain}
\newtheorem{thm}{Theorem}

\newtheorem{prop}{Proposition}

\theoremstyle{definition}
\newtheorem{defn}{Definition}
\newtheorem{disc}{Discussion}
\newtheorem{note}{Note}
\newtheorem{remk}{Remark}

\usepackage[scr = dutchcal]{mathalfa}

\usepackage{graphicx,stackengine} %
\usepackage{relsize}

\usepackage[ruled]{algorithm2e}
\usepackage{algcompatible}

\usepackage{lipsum}
\usepackage[caption=false, font=normalsize, labelfont=sf, textfont=sf]{subfig}

\usepackage{rotating}

\begin{document}

\begin{frontmatter}

\title{Stationarity of Time-Series on Graph via Bivariate Translation Invariance}



\author[mymainaddress1]{Amin~Jalili}
\author[mymainaddress2]{Chong-Yung Chi}
\address{Institute of Communications Engineering, National Tsing Hua University}
\fntext[mymainaddress1]{Corresponding author, Student member, IEEE, e-mail: amin.jalili@ieee.org.}
\fntext[mymainaddress2]{Fellow, IEEE, e-mail: cychi@ee.nthu.edu.tw}

\fntext[]{This work was supported by the Ministry of Science and Technology, R.O.C., under Grant MOST 107-2221-E-007-021, MOST 108-2634-F-007-004 and MOST 108-2221-E-007-024. It is noted that pre-print of this paper is posted on \texttt{arXiv.org}~\cite{Amin2020stationarity}.}
%

\address{Hsinchu, Taiwan}

\begin{abstract}
	Stationarity is a cornerstone in classical signal processing (CSP) for modeling and characterizing various stochastic signals for the ensuing analysis. However, in many complex real world scenarios, where the stochastic process lies over an irregular graph structure, CSP discards the underlying structure in analyzing such structured data. Then it is essential to establish a new framework to analyze the high-dimensional graph structured stochastic signals by taking the underlying structure into account. To this end, looking through the lens of operator theory, we first propose a new \textit{bivariate isometric joint translation operator (JTO)} consistent with the structural characteristic of translation operators in other signal domains. Moreover, we characterize time-vertex filtering based on the proposed JTO. Thereupon, we put forth a new definition of joint wide-sense stationary (JWSS) signals in time-vertex domain using the proposed isometric JTO together with its spectral characterization. Then a new joint power spectral density (JPSD) estimator, called generalized Welch method (GWM), is presented. Simulation results are provided to show the efficacy of this JPSD estimator. Furthermore, to show the usefulness of JWSS modeling, we focus on the \textit{classification of time-series on graph}. To that end, by modeling the brain Electroencephalography (EEG) signals as JWSS processes, we use JPSD as the feature for the \textit{Emotion} and \textit{Alzheimer's disease (AD) recognition}. Experimental results demonstrate that JPSD yields superior Emotion and AD recognition accuracy in comparison with the classical power spectral density (PSD) and graph PSD (GPSD) as the feature set for both applications. Eventually, we provide some concluding remarks. 
\end{abstract}

\begin{keyword}
	Graph signal processing; joint translation operator; joint power spectral density; joint wide-sense stationarity. 
\end{keyword}

\end{frontmatter}


\section{Introduction}

Beyond doubt, we are in the era of big data in which massive amount of information is generated at a fast pace and this poses new challenges for the data science. Often, the \textit{big structured data} lies over an irregular structure, but the classical signal processing (CSP) disregards the underlying \textit{topological structure}. Graphs, as the powerful mathematical tool, have been widely used in biology, physics, and computer science to model the underlying complex structure of non-Euclidean data through pairwise relations between objects. Connecting the concepts from algebraic graph theory to the CSP gave birth to the field of graph signal processing (GSP)~\cite{Shuman2013Emerging,Chung1997Spectral,HAMMOND2011129Wavelets,Sandryhaila2013Discrete,Puschel2008Algebraic,STANKOVIC2020Vertex} as a theoretical discipline for analyzing the structured data. In the recent years, GSP emerged via numerous theoretical research works for tackling challenging problems in modern signal processing and data science~\cite{Shuman2016Vertex,perraudin2017graph,Bronstein2017Geometric,perraudin2018Global,Giannakis2018Topology,Dong2019Learning,Dong2014Multi,Grassi2018Time}. In particular,~\cite{Grassi2018Time} proposed a new framework for analyzing time-varying graph signals through a meaningful representations of time-series on graph. Stationarity and its important subclass wide-sense stationarity play an essential role in statistical signal processing and time-series analysis. By the classical definition, a signal is temporal wide-sense stationary (TWSS) if \textit{its mean and autocorrelation functions are translation invariant}. Likewise, this concept is of paramount importance concerning time-series on irregular graph structures.

Stationarity on graph is first defined by  Girault \textit{et al.}~\cite{Girault2015Stationary,Girault2015Trans} via isometric graph translation operator. Afterwards, Perraudin and Vandergheynst~\cite{Perraudin2017Stationary} defined stationarity on graph via a novel localization operator. Moreover, Marques~\textit{et~al.}~\cite{Marques2017Stationary} defined weak stationarity of random graph signals using the adjacency matrix (or graph Laplacian matrix) as the graph shift operator. These approaches lead to \textit{almost the same definition of stationarity on graph} in spite of their different initial ideas~\cite{Loukas2017JStationary}. Then, Segarra~\textit{et~al.}~\cite{SEGARRA2018Statistical} defined stationarity in time-vertex domain based on the weighted adjacency matrix of joint/product graph. By generalizing a classical ``\textit{filtering interpretation}" of stationarity from Euclidean space, Loukas and Perraudin~\cite{Perraudin2017Towards} defined joint wide-sense stationary (JWSS) signals via time-vertex filtering. Noticeably, they have shown that stationarity in the time-vertex domain is more general than defining the stationarity on joint/product graph. Using this general notion of joint wide-sense stationarity, Isufi~\textit{et al.}~\cite{Isufi2019Forecasting} extended the classical vector autoregressive and vector autoregressive moving average recursions for modeling and forecasting time-series on graph.

{
In this paper, a novel approach, beyond~\cite{SEGARRA2018Statistical,Perraudin2017Towards,Isufi2019Forecasting}, is proposed for characterizing the stationarity of time-series on graph via bivariate translation invariance in the time-vertex domain. First, we design a bivariate isometric joint translation operator (JTO) in the time-vertex domain. Then we put forth a new definition of wide-sense stationarity of time-series on graph based on the proposed JTO followed by its spectral characterization. Next, the joint power spectral density (JPSD) estimation of JWSS processes is proposed using a generalized Welch method (GWM) followed by some simulations to show its effectiveness. Finally, we demonstrate the applicability of proposed framework for the classification of time-series on irregular graphs by providing experimental results concerning Emotion and Alzheimer's disease (AD) recognition from Electroencephalography (EEG) data.

\raggedright
\justify
{${{\mathbf{Notations.}}}$}
Matrices and vectors are denoted by uppercase and lowercase boldface letters, $\bm{A}$ and $\bm{a}$, respectively. The $n$-th element of a vector is indexed by $\bm{a}[n]$, and the entry in $n$-th row and $m$-th column of a matrix is denoted by $\bm{A}[n,m]$. $\mathbf{R} = [{{\bm{\mathsf{R}}}_{\left(m,n\right)}}]$ is a block matrix where ${{\bm{\mathsf{R}}}_{\left(m,n\right)}}$ is its submatrix in the $m$-th row and $n$-th column partition. Other notations are as follows: $\bm{A}^{\mathsf T}$, $\widebar{\bm{A}}$, and $\bm{A}^{*} = (\widebar{\bm{A}})^{\mathsf T}$ stand for the transpose, conjugate, and complex conjugate transpose of the matrix $\bm{A}$, respectively. Moreover, ${\mathsf{vec}}(\bm{A})$ stands for the column vector by stacking all the columns of $\bm{A}$ sequentially, ${\mathsf{Diag}}(\bm{a})$ represents a diagonal matrix by placing the elements of vector $\bm{a}$ on the main diagonal, and ${\mathsf{Diag}}(\bm{A})$ is equivalent to ${\mathsf{Diag}}({\mathsf{vec}}(\bm{A}))$, while ${\mathsf{diag}}(\bm{A})$ represents the column vector containing the diagonal elements of matrix $\bm{A}$. We use $\|\bm{a}\|_p$ and $|a|$ as the $p$-norm of $\bm{a}$ and absolute value of $a$, respectively. On the other hand, $|\bm{A}|$ is a matrix with $(n,m)$-th element equal to $|\bm{A}[n,m]|$. $\mathsf{row}_{k}(\bm{A})$ stands for the $k$-th row of matrix $\bm{A}$. Also, $\bf{I}$, $\bf{0}$, and $\bf{1}$ denote the identity matrix, matrix/column vector of all zeros, and matrix/column vector of all ones (their dimensions may be indicated by their subscript for some emphasized cases), respectively. Symbols $\otimes$ and $\odot$ represent the Kronecker product and Hadamard (element-wise) product, respectively. Besides, $\mi = \sqrt{-1}$ and $\llbracket a,b \rrbracket$ represents the set of integers between $a$ and $b$ inclusive. The Kronecker sum, denoted by $\oplus$, is defined as: $\bm{A}\oplus \bm{B} \coloneqq \bm{A}\otimes \mathbf{I}_{M} + \mathbf{I}_{N}\otimes \bm{B}$ where $\bm{A} \in \mathbb{C}^{N\times N}$ and $\bm{B} \in \mathbb{C}^{M\times M}$. On the other hand, the direct sum of a set of matrices $\bm{\mathsf{A}}_i$ for $i \in \llbracket 1,d \rrbracket$ is defined as $\widehat\oplus_{i=1}^d\bm{\mathsf{A}}_i \coloneqq {\mathsf{Diag}}(\bm{\mathsf{A}}_1, \bm{\mathsf{A}}_2, \ldots, \bm{\mathsf{A}}_d)$. Then $\mathbb{C}^{N\times N}$ ($\mathbb{R}^{N\times N}$) is the set of $N\times N$ complex (real) matrices. Besides, $\mathbb{C}^{N}$ ($\mathbb{R}^{N}$) is the set of $N\times 1$ complex (real) vectors. $\mathbb{R}_{+}$ ($\mathbb{Z}_{+}$) accounts for the set of nonnegative real (integer) numbers. Note that $\mathbb{E}[\cdot]$ accounts for the statistical expectation. Moreover, $\mathsf C$, $\mathsf D$, $\mathsf G$, and $\mathsf J$ represent the continuous-time, discrete-time, graph, and joint time-vertex domains, respectively.

\begin{note}\label{20200621_1849}
	Let $\mathbf{L}$ be an $N\times N$ Hermitian matrix with eigenvalue decomposition $\mathbf{L} = \mathbf{\Phi} \mathbf{\Lambda} \mathbf{\Phi}^{*}$ where $\mathbf{\Phi}$ is a unitary eigenbasis matrix $\mathbf{\Phi}\mathbf{\Phi}^{*} = \mathbf{\Phi}^{*}\mathbf{\Phi} = \mathbf{I}$ and ${\mathbf{\Lambda}} = {\mathsf{Diag}}\left(\left[\lambda_{0}, \ldots, \lambda_{N-1}\right]\right)$ is the diagonal eigenvalue matrix. Then, the univariate matrix function is defined as $h(\mathbf{L}) \coloneqq \mathbf{\Phi} h(\mathbf{\Lambda}) \mathbf{\Phi}^{*}$ where $h(\mathbf{\Lambda}) = {\mathsf{Diag}}([h(\lambda_{0}), \ldots, h(\lambda_{N-1})])$ in which $h: \mathbb{R} \to \mathbb{C}$~\cite{Higham2008Functions}. Note that $h$ can also be a multivariate matrix function. 
\end{note}

\section{Background}
\label{secBackground}

\raggedright
\justify
{${\mathsf{\mathbf{Vertex~Harmonic~Analysis.}}}$}
Let ${\mathsf G} \coloneqq (\mathscr{V}, \mathscr{E}, W)$ denote a fixed graph with finite vertex set $\mathscr V$ with the cardinality $|\mathscr{V}| = N$, $\mathscr{E} = \{(n_{1},n_{2}) | \; n_{1},n_{2} \in \mathscr{V}, n_{1} \sim n_{2}\} \subseteq \mathscr{V} \times \mathscr{V}$ is the edge set and $W: \mathscr{V} \times \mathscr{V} \rightarrow \mathbb{R}_{+}$ is a weight function. This function yields the weighted adjacency matrix as ${\mathbf W}_{\mathsf G} = [w_{n_{1},n_{2}}] \in \mathbb{R}_{+}^{N \times N}$ where $w_{n_{1},n_{2}}$ represents the strength of the connection between nodes $n_{1}$ and $n_{2}$. 
The neighborhood of vertex $v_{n}$ is the set of vertices connected to $v_{n}$ by an edge.
Throughout this paper, we assume that the graph is finite, weighted, connected, and undirected.
A graph signal, represented in a compact form by the vector $\bm{f} \in \mathbb{C}^{N}$, is defined by the function $f_{\mathsf G}: \mathscr{V} \rightarrow \mathbb{C}$ where $\bm{f}[n]$ is the function value at the vertex $n$. Then the graph Laplacian matrix is defined as 
$
{\mathbf L}_{\mathsf G} \coloneqq \mathbf {\mathsf{Diag}}\left({\mathbf W}_{\mathsf G}{\mathbf 1}\right) - {\mathbf W}_{\mathsf G}.
$
Since $\mathbf{L}_{\mathsf G}$ is a symmetric positive semi-definite matrix, the eigenvalue decomposition (EVD) of $\mathbf{L}_{\mathsf G}$ can be written as 
\begin{align}\label{20200620_0830}
{\mathbf{L}_{\mathsf G}} = {\mathbf{\Phi}_{\mathsf G}}{\mathbf{\Lambda}_{\mathsf G}} \mathbf{\Phi}_{\mathsf G}^{*},
\end{align}
where $\mathbf{\Phi}_{\mathsf G} = [{\bm{\varphi}}_{{\mathsf G},0}, \ldots, {\bm{\varphi}}_{{\mathsf G}, N-1}] \in \mathbb{C}^{N\times N}$ is a unitary matrix consisting of a complete set of orthonormal eigenvectors $\mathscr{B}_{\mathsf G} \coloneqq \left\{\bm{\varphi}_{{\mathsf G},\ell}, \forall \ell \in {\llbracket 0,N-1 \rrbracket}\right\}$, such that ${\bm{\varphi}}_{{\mathsf G},\ell}$ is equal to the \textit{$\ell$-th graph Fourier mode}~\cite{Shuman2013Emerging}, and 
\begin{align}\label{20200616_1648} 
{\mathbf{\Lambda}}_{\mathsf G} = {\mathsf{Diag}}\left(\left[\lambda_{{\mathsf G},0}, \ldots, \lambda_{{\mathsf G}, N-1}\right]\right),
\end{align}
contains the eigenvalues of ${\mathbf L}_{\mathsf G}$ where without loss of generality, we assume: $0=\lambda_{{\mathsf G},0} < \lambda_{{\mathsf G},1} \leq \cdots \leq \lambda_{{\mathsf G},N-1} = \lambda_{{\mathsf G},\max}$. 
Shuman~\textit{et al.}~\cite{Shuman2013Emerging} specified that $\lambda_{\ell}$ for $\ell \in {\llbracket 0,N-1 \rrbracket}$ carries the frequency notion in graph setting.
The reduced graph frequencies (one may consider as the angular frequencies on graph) is defined as~\cite{Girault2015Trans}
\begin{align}\label{20200616_1645}
\bm{\mathscr{W}}_{\mathsf{G}} \coloneqq \big\{\omega_{{\mathsf{G}},\ell} \coloneqq \pi \sqrt{\lambda_{{\mathsf G},\ell}/\rho_{\mathsf{G}}}, \quad \ell \in {\llbracket 0,N-1 \rrbracket}\big\},
\end{align} 
where $\rho_{\mathsf{G}}$ is an upper bound of $\lambda_{\mathsf{G},\ell}$ defined as
\begin{align}\label{roCoeff}
\rho_{\mathsf{G}} \coloneqq \max_{i\in\mathscr{V}} \sqrt{2d_{i}(d_{i}+d_{i}^{\prime})}, \quad d_{i}^{\prime}\coloneqq \frac{\sum_{j=1}^{N} w_{ij}d_{j}} {d_{i}},
\end{align}
where $d_{i} = \sum_{j \in \mathcal{N}_{i}} w_{i,j}$. On the other hand, the graph Fourier transform (GFT) and its inverse can be expressed as~\cite{Shuman2013Emerging} 
\begin{align}\label{2020-2-6-0700}
{\textrm{GFT:}}\; \widehat{\bm{x}} = \mathcal{F}_{\mathsf G}\bm{x} =\mathbf{\Phi}_{\mathsf G}^{*} \bm{x} \;\; \longleftrightarrow \;\;
{\textrm{IGFT:}}\; \bm{x} = \mathcal{F}_{\mathsf G}^{-1}\widehat{\bm{x}} = \mathbf{\Phi}_{\mathsf G} \widehat{\bm{x}},
\end{align}
where $\mathcal{F}_{\mathsf G}$ and $\mathcal{F}_{\mathsf G}^{-1}$ account for the GFT and inverse GFT (IGFT) operator, respectively.

\raggedright
\justify
{${{\mathbf{Discrete\!-\!Time~Harmonic~Analysis.}}}$}
Let $x(n), \; n\in {\llbracket 1,M \rrbracket}$ be a discrete-time signal of finite length $M$. The discrete Fourier transform (DFT) operator $\mathcal{F}_{\mathsf D}$ and its inverse $\mathcal{F}_{\mathsf D}^{-1}$ can be represented in a matrix form as~\cite{vetterli2014Foundations}
\begin{align}\label{2020-2-6-0705}
{\textrm{DFT:}}\; \widehat{\mathbf{x}} = \mathcal{F}_{\mathsf D}{\mathbf{x}} = \mathbf{\Phi}^{*}_{\mathsf D}  {\mathbf{x}} \;\; \longleftrightarrow \;\;
{\textrm{IDFT:}}\; {\mathbf{x}} = \mathcal{F}_{\mathsf D}^{-1}{\widehat{\mathbf{x}}} = {\mathbf{\Phi}_{\mathsf D}}\widehat{\mathbf{x}},
\end{align}
where ${\mathbf x} \coloneqq \left[x(1), x(2), \ldots, x(M)\right]^{\mathsf T}$ is the discrete-time signal in vector form. Here, $\mathbf{\Phi}_{\mathsf D}^{*}$ and $\mathbf{\Phi}_{\mathsf D}$ account for the DFT and inverse DFT (IDFT) matrices, respectively, such that
$$
\mathbf{\Phi}_{\mathsf D}[n+1,k+1] = \frac{1}{\sqrt{M}}\exp({{\mi n\omega_{{\mathsf D}, k}}}), \quad n,k \in {\llbracket 0,M-1 \rrbracket},
$$
and the set of angular frequencies is defined as 
\begin{align}\label{20200627_0537}
\bm{\mathscr{W}}_{\mathsf{D}} \coloneqq \{\omega_{{\mathsf D}, k} \coloneqq 2\pi k/M, \; k\in {\llbracket 0,M-1 \rrbracket}\}.
\end{align}
By a classic interpretation, discrete-time domain can be modeled as a $M$-Cycle graph $\mathsf D$ with all edge weights equal to unity. Moreover, the symmetric time Laplacian matrix $\mathbf{L}_{\mathsf D}$, corresponding to the graph $\mathsf{D}$, is defined as the second-order derivative in discrete-time domain up to a negative sign and, as a circulant matrix, it can be diagonalized as~\cite{Grassi2018Time}
\begin{align}\label{20200207_0945}
{\mathbf{L}_{\mathsf D}} = {\mathbf{\Phi}_{\mathsf D}}{\mathbf{\Lambda}_{\mathsf D}} \mathbf{\Phi}_{\mathsf D}^{*},
\end{align} 
where the corresponding eigenvalue matrix is
\begin{align}\label{20200617_0826}
{\mathbf{\Lambda}}_{\mathsf D} = {\mathsf{Diag}}\left(\left[\lambda_{{\mathsf D},0}, \ldots, \lambda_{{\mathsf D}, M-1}\right]\right),
\end{align}
such that $\lambda_{{\mathsf D}, k} = 2(1-\cos(\omega_{{\mathsf D}, k}))$ for all $k \in {\llbracket 0,M-1 \rrbracket}$.

\raggedright
\justify
{${\mathsf{\mathbf{Joint~Time\!-\!Vertex~Harmonic~Analysis.}}}$}
A time-varying graph signal is represented in a compact form by the matrix 
$
{\bm{X}} = [{\bm{x}}_{1}, {\bm{x}}_{2}, \ldots,{\bm{x}}_{M}] \in \mathbb{R}^{N \times M},
$
where ${\bm{x}}_{k}$ denotes the graph signal at discrete-time $k \in {\llbracket 1,M \rrbracket}$ with a fixed sampling interval. Then the joint time-vertex Fourier transform (JFT) is defined as
$
\widehat{{\bm{X}}} \coloneqq \mathbf{\Phi}_{\mathsf G}^{*} {\bm{X}} {\widebar{\mathbf{\Phi}}_{\mathsf D}}
$
where $\mathbf{\Phi}_{\mathsf G}^*$ and $\mathbf{\Phi}_{\mathsf D}$ are the GFT and IDFT matrices, respectively~\cite{Loukas2016Frequency,Grassi2018Time}.  
The two-dimensional spectrum of time-vertex signal $\bm{X}$, namely joint spectrum, can be described as
\begin{align}\label{20200606_1231}
{\mathbf{\Omega}} \coloneqq [(\omega_{\mathsf{G},\ell},\omega_{\mathsf{D},k})], \quad \ell \in {\llbracket 0,N-1 \rrbracket},\; k \in {\llbracket 0,M-1 \rrbracket},
\end{align}
containing frequency pairs $(\omega_{\mathsf{G},\ell},\omega_{\mathsf{D},k})$ where $\omega_{\mathsf{G},\ell}$ and $\omega_{\mathsf{D},k}$ are the $\ell$-th and $k$-th angular frequencies on graphs $\mathsf{G}$ and $\mathsf{D}$, respectively.
In a compact form, JFT and its inverse can be rewritten as~\cite{Loukas2016Frequency} 
\begin{align}\label{2020-2-6-07010}
{\textrm{JFT:}}\; \widehat{\bm{x}} = \mathcal{F}_{\mathsf J}\bm{x} =\mathbf{\Phi}_{\mathsf J}^{*} \bm{x} \;\; \longleftrightarrow \;\;
{\textrm{IJFT:}}\; \bm{x} = \mathcal{F}_{\mathsf J}^{-1}\widehat{\bm{x}} = \mathbf{\Phi}_{\mathsf J} \widehat{\bm{x}},
\end{align}
where ${\bm{x}} = {\mathsf{vec}}({\bm{X})}$ is called the ``time-vertex signal" and 
\begin{align}\label{JFTMatrix1655}
{\mathbf{\Phi}_{\mathsf J}} \coloneqq {\mathbf{\Phi}_{\mathsf D}} \otimes {\mathbf{\Phi}_{\mathsf G}} \in \mathbb{C}^{NM\times NM},
\end{align}
is a \textit{unitary matrix}. This consists of a complete set of orthonormal eigenvectors $$\bm{\mathscr{B}}_{\mathsf J} \coloneqq \left\{\bm{\varphi}_{{\mathsf J},j}, \forall j \in {\llbracket 0,NM-1 \rrbracket}\right\},$$ such that ${\bm{\varphi}}_{{\mathsf J},\ell}$ is equivalent to the \textit{$\ell$-th joint Fourier mode}. One can easily see that $\boldsymbol{\varphi}_{\mathsf{J},0} = \mathbf{1}_{NM}/\sqrt{NM}$.
Likewise the graph setting and discrete-time domain, joint time-vertex domain can also be modeled by a \textit{joint graph} (or product graph) which is equal to the Cartesian product of undirected graph $\mathsf G$ and $M$-Cycle graph $\mathsf D$ (see~\cite[Figure 2]{Loukas2016Frequency}). Moreover, the joint Laplacian matrix $\mathbf{L}_{\mathsf J}$, corresponding to the joint graph, can be described as~\cite{Grassi2018Time}
\begin{align}\label{20200606_0944}
\mathbf{L}_{\mathsf J} = \mathbf{L}_{\mathsf D} \oplus \mathbf{L}_{\mathsf G} = {\mathbf{\Phi}_{\mathsf J}}{\mathbf{\Lambda}_{\mathsf J}} \mathbf{\Phi}_{\mathsf J}^{*},
\end{align}
where the corresponding eigenvalue matrix is
\begin{align}\label{20200616_1657}
{\mathbf{\Lambda}}_{\mathsf J} = {\mathbf{\Lambda}}_{\mathsf D} \oplus {\mathbf{\Lambda}}_{\mathsf G} = {\mathsf{Diag}}\left(\left[\lambda_{{\mathsf J},0}, \ldots, \lambda_{{\mathsf J}, NM-1}\right]\right),
\end{align}
contains the eigenvalues of ${\mathbf L}_{\mathsf J}$ corresponding to the sum of all eigenvalue pairs of $\mathbf{L}_{\mathsf G}$ and $\mathbf{L}_{\mathsf D}$ (cf.~\eqref{20200616_1648},~\eqref{20200617_0826}).
One can write the $\ell$-th eigenvalue of ${\mathbf L}_{\mathsf J}$ as follows
\begin{align}\label{20200616_1658}
\lambda_{{\mathsf J},j} = \lambda_{{\mathsf G},\ell} + \lambda_{{\mathsf D},k}, \quad j \in {\llbracket 0,NM-1 \rrbracket},
\end{align}
where 
\begin{align}\label{20200617_0929}
j = \ell + kN, \quad \ell \in {\llbracket 0,N-1 \rrbracket}, \, k \in {\llbracket 0,M-1 \rrbracket}.
\end{align}


\section{Joint Time-Vertex Translation Operator}

First, we briefly discuss the translation operator in continuous-time, discrete-time, and graph domains followed by designing bivariate isometric translation operator in time-vertex domain.  

\begin{remk}\label{20200614_1227}
	Let $x(t)$ be a continuous-time signal and $\widehat{x}(\xi) = (\mathcal{F}_{\mathsf C} x)(t)$ be its Fourier transform where $\mathcal{F}_{\mathsf C}$ is the Fourier transform operator. The continuous translation operator is defined as $(\mathcal{T}_{\mathsf C}^{\tau}x)(t) = x(t-\tau)$ with $\tau$ as the translation value. This operator can be formulated in the abstract form as
	\begin{align}\label{shiftOpContiDomain}
	\mathcal{T}_{\mathsf C}^{\tau} = \mathcal{F}_{\mathsf C}^{-1}\mathcal{P}_{\mathsf C}^{\mathsf \tau}\mathcal{F}_{\mathsf C},
	\end{align}
	where $\mathcal{P}_{\mathsf C}^{\tau} = \exp{(-2\pi\mi\tau\xi )}$ is the linear phase shift operator. 
	
	\vspace*{0.1cm}
	\textit{Proof:}
	{\normalfont
	See~\ref{AppendixI}.
	}
	\hfill $\openbox$ 
\end{remk}

\begin{remk}
	Let $x(n)$ be a discrete-time signal where $n\in {\llbracket 1,M \rrbracket}$ and $\mathcal{T}_{\mathsf D}^{\upsilon}$ be the right-circular translation operator in discrete-time domain with the translation value $\upsilon$ defined as $(\mathcal{T}_{\mathsf D}^{\upsilon}x)(n) \coloneqq x(n-\upsilon)$. Let ${\mathbf x}\coloneqq \left[x(1), x(2), \ldots, x(M)\right]^{\mathsf T}$ be the signal in vector form. Then, $\mathbf{T}_{\mathsf D}\mathbf{x} = [\mathbf{e}_2, \mathbf{e}_3, \ldots, \mathbf{e}_{M}, \mathbf{e}_{1}] \mathbf{x}$ where $\mathbf{T}_{\mathsf D}$ is the matrix representation of $\mathcal{T}_{\mathsf D}$ and $\mathbf{e}_i$ is the $M\times 1$ unit vector with the $i$-th entry equal to 1.
	The discrete-time translation operator can be unitarily diagonalized by the DFT matrix as
	\begin{align}\label{DTOFullrepres0907}
	\mathbf{T}_{\mathsf D} = {\mathbf{\Phi}_{\mathsf D}} {{\mathbf{P}}}_{\mathsf D} \mathbf{\Phi}_{\mathsf D}^{*},
	\end{align}
	where ${{\mathbf{P}}}_{\mathsf D} \coloneqq \exp(-\mi {\bm{\mathbf{M}}}_{\mathsf D})$ such that
	\begin{align}\label{20200624_1838}
	{\bm{\mathbf{M}}}_{\mathsf D} \coloneqq {\mathsf{Diag}}\big([\omega_{{\mathsf D}, 0},\ldots, \omega_{{\mathsf D}, M-1}]\big)
	\end{align}
	is the angular frequency matrix in discrete-time domain. 
	\hfill $\openbox$
\end{remk}

\begin{defn}(\textit{Graph Shift Operator}~\cite{Girault2015Trans})\label{gto}
	The isometric graph shift operator on graph $\mathsf{G}$ is defined as 
	\begin{align}\label{graphTransOp}
	\mathbf{T}_\mathsf{G} \coloneqq
	\exp\big(-\mi\pi\sqrt{\mathbf{L}_{\mathsf{G}} /\rho_{\mathsf{G}}}\big). \qquad (\text{cf.}~\eqref{roCoeff})
	\end{align}
\end{defn}
The isometric graph shift operator can be written as
\begin{align}\label{20200617_0828}
\mathbf{T}_{\mathsf G} = \mathbf{\Phi}_{\mathsf G}  
\mathbf{P}_{\mathsf G} \mathbf{\Phi}_{\mathsf G}^{*},
\end{align}
where $\mathbf{P}_{\mathsf G} = \exp{\left(-\mi {{\mathbf{M}}}_{\mathsf G}\right)}$ such that
\begin{align}\label{20200624_1837}
\bm{\mathbf{M}}_{\mathsf G} \coloneqq {\mathsf{Diag}}\left(\left[\omega_{{\mathsf{G}},0}, \ldots, \omega_{{\mathsf{G}}, N-1}\right]\right) \quad \textrm{(cf.~\eqref{20200616_1645})}
\end{align}
is a diagonal matrix containing the reduced graph frequencies, and $\mathbf{\Phi}_{\mathsf G}^{*}$ is the GFT matrix.

\begin{defn}(\textit{Joint Translation Operator})\label{JTVtranOper}
	Let $\mathbf{T}_{\mathsf D}$ and $\mathbf{T}_{\mathsf G}$ be the translation operators in discrete-time and graph domains, respectively. We define $\mathcal{T}_{\mathsf J}^{(\upsilon,\vartheta)}: \mathbb{R}^{N\times M} \to \mathbb{C}^{N\times M}$ as the $(\upsilon,\vartheta)$-translation of time-varying graph signal ${\bm{X}}$ as follows
	\begin{align}\label{20200921_2033}
	{\bm{X}}^{(\upsilon,\vartheta)} = \mathcal{T}_{\mathsf J}^{(\upsilon,\vartheta)}  {\bm{X}} \coloneqq \mathbf{T}_{\mathsf G}^{\vartheta} {\bm{X}} \left(\mathbf{T}_{\mathsf D}^{\mathsf T}\right)^{\upsilon}, 
	\end{align}
	where $\vartheta, \upsilon \in \mathbb{Z}_{+}$ account for the translation value in graph and discrete-time domains, respectively. 
\end{defn}

\begin{remk}
	To represent the bivariate joint translation operation in matrix-vector multiplication form, one can write
	\begin{align}\label{20200615_1342}
	{\bm{x}}^{(\upsilon,\vartheta)} = \mathbf{T}_{\mathsf J}^{(\upsilon,\vartheta)}{\bm{x}},
	\end{align}
	where ${\bm{x}}^{(\upsilon,\vartheta)} = {\mathsf{vec}}\big({\bm{X}}^{(\upsilon,\vartheta)}\big)$, ${\bm{x}} = {\mathsf{vec}}({\bm{X}})$, and thanks to the nice property of Kronecker product, the matrix representation of bivariate isometric JTO can be obtained as
	\begin{align}\label{20200624_1100}
	\mathbf{T}_{\mathsf J}^{(\upsilon,\vartheta)} = \mathbf{T}_{\mathsf D}^{\upsilon} \otimes \mathbf{T}_{\mathsf G}^{\vartheta}.
	\end{align}
	It should be underlined that the proposed bivariate JTO is genuinely defined in the \textit{time-vertex domain} as a two-dimensional operation on the matrix ${\bm{X}}$. However, for notational simplicity, in the ensuing presentation, we may use the compact form of this operation given by~\eqref{20200615_1342} instead.
	\hfill $\openbox$
\end{remk}

\begin{prop}\label{20200622_1443}
	\normalfont
	The JTO $\mathcal{T}_{\mathsf J}^{(\upsilon,\vartheta)}$ (cf. Definition~\ref{JTVtranOper}) is a unitary operator, and hence isometric.
	
	\vspace*{0.1cm}
	\textit{Proof:} 
	\normalfont{
		It is sufficient to prove it for the unit joint time-vertex translation simply denoted by $\mathbf{T}_{\mathsf J}$. Then we have
		\begin{align*}
		\begin{split}
		\mathbf{T}_{\mathsf J}\mathbf{T}_{\mathsf J}^{*} 
		= \left(\mathbf{T}_{\mathsf D} \otimes \mathbf{T}_{\mathsf G}\right) \left(\mathbf{T}_{\mathsf D} \otimes \mathbf{T}_{\mathsf G}\right)^{*} 
		= \left(\mathbf{T}_{\mathsf D} \mathbf{T}_{\mathsf D}^{*}\right) \otimes \left(\mathbf{T}_{\mathsf G} \mathbf{T}_{\mathsf G}^{*}\right)
		= \mathbf{I}_{M} \otimes \mathbf{I}_{N} = \mathbf{I}_{NM},
		\end{split}
		\end{align*}
		where the third equality holds since $\mathbf{T}_{\mathsf G}$ and $\mathbf{T}_{\mathsf D}$ are unitary matrices. Similarly, it can be shown that $\mathbf{T}_{\mathsf J}^{*} \mathbf{T}_{\mathsf J} = \mathbf{I}_{NM}$. \hfill $\blacksquare$}
\end{prop}

\begin{sidewaystable}
	\caption{Structural characteristic of Translation operators in various signal domains} 
	\vspace*{0.1cm}
	\ra{1.3}
	\label{tab:tbl1}
	\centering 
	\small
		\begin{tabular}{@{}l l l@{}} 
		\toprule[1pt]
		\textbf{Domain} & \textbf{Structural characteriestic} & \textbf{Description}   \\ 
		\hline
		\multirow{2}{*}{Continuous-time (cf.~\eqref{shiftOpContiDomain})}  & \multirow{2}{*}{$
		\mathcal{T}_{\mathsf C}^{\tau} = \mathcal{F}_{\mathsf C}^{-1}\mathcal{P}_{\mathsf C}^{\mathsf \tau}\mathcal{F}_{\mathsf C}
		$}	
		& $\mathcal{F}_{\mathsf C}$: Continuous-time Fourier transform operator, $\mathcal{P}_{\mathsf C}^{\tau} = \exp{\left(-\mi 2\pi \xi \tau\right)}$,\\ 
		& & $\mathcal{M}_{\mathsf C}$: Angular frequency multiplication operator --- $\mathcal{M}_{\mathsf C}\widehat{x}(\xi) \coloneqq 2\pi \xi\widehat{x}(\xi)$
		\\
		\hline
		\multirow{2}{*}{Discrete-time (cf.~\eqref{DTOFullrepres0907})}  & \multirow{2}{*}{$
		\mathbf{T}_{\mathsf D}^{\upsilon} = \mathbf{\Phi}_{\mathsf D} \mathbf{P}_{\mathsf D}^{\upsilon} \mathbf{\Phi}_{\mathsf D}^{*}
		$}	
		& $\mathbf{\Phi}_{\mathsf D}^{*}$: DFT matrix, $\mathbf{P}_{\mathsf D}^{\upsilon} \coloneqq \exp\big({-\mi\upsilon{\bm{\mathbf{M}}}_{\mathsf D}}\big)$, \\ 
		& & ${{\mathbf{M}}}_{\mathsf D}$: Diagonal matrix of discrete angular frequencies  
		\\
		\hline
		\multirow{2}{*}{Graph (cf.~\eqref{20200617_0828})}   & \multirow{2}{*}{$
		\mathbf{T}_{\mathsf G}^{\vartheta} = \mathbf{\Phi}_{\mathsf G} \mathbf{P}_{\mathsf G}^{\vartheta} \mathbf{\Phi}_{\mathsf G}^{*}
		$}	
		& $\mathbf{\Phi}_{\mathsf G}^{*}$: GFT matrix, $\mathbf{P}_{\mathsf G}^\vartheta \coloneqq \exp{\big(-\mi \vartheta {\mathbf{M}}_{\mathsf G}\big)}$, \\ 
		& & ${\mathbf{M}}_{\mathsf G}$: Diagonal matrix of angular frequencies in graph setting 
		\\
		\hline
		\multirow{2}{*}{Joint time-vertex (cf.~\eqref{JTVtransAbstract})}  & \multirow{2}{*}{$
		\mathbf{T}_{\mathsf J}^{(\upsilon,\vartheta)} = \mathbf{\Phi}_{\mathsf J} \mathbf{P}_{\mathsf J}^{(\upsilon,\vartheta)} \mathbf{\Phi}_{\mathsf J}^{*}
		$}	
		& $\mathbf{\Phi}_{\mathsf J}^{*}$: JFT matrix, $\mathbf{P}_{\mathsf J}^{(\upsilon,\vartheta)} \coloneqq \exp\big(-\mi {\bm{\mathbf{M}}}_{\mathsf J}^{(\upsilon,\vartheta)}\big)$, \\ 
		& & ${\bm{\mathbf{M}}}_{\mathsf J}^{(\upsilon,\vartheta)}$: Diagonal matrix of joint angular frequencies 
		\\
		\bottomrule[1pt]
		\end{tabular}
\end{sidewaystable}


\begin{thm}\label{theoremJTVtrans}
	The proposed JTO $\mathcal{T}_{\mathsf J}^{(\upsilon,\vartheta)}$ can be written in matrix form as follows
	\begin{align}\label{JTVtransAbstract}
	\mathbf{T}_{\mathsf J}^{(\upsilon,\vartheta)} = \mathbf{\Phi}_{\mathsf J} \mathbf{P}_{\mathsf J}^{(\upsilon,\vartheta)} \mathbf{\Phi}_{\mathsf J}^{*}, \qquad \forall \upsilon, \vartheta \in \mathbb{Z}_{+}, 
	\end{align}
	where the bivariate phase shift operator is
	\begin{align}\label{20200611_1617}
	\mathbf{P}_{\mathsf J}^{(\upsilon,\vartheta)} = \exp(-\mi \bm{\mathbf{M}}_{\mathsf J}^{(\upsilon,\vartheta)}),
	\end{align}
	and 
	$
	\bm{\mathbf{M}}_{\mathsf J}^{(\upsilon,\vartheta)} = {\mathsf{Diag}}({\boldsymbol\xi}^{\left(\upsilon,\vartheta\right)})
	$
	such that 
	\begin{align}\label{jointFreq0832}
	{\boldsymbol\xi}^{(\upsilon,\vartheta)} 
	= 
	{\mathsf{vec}}\left(
	\begin{bmatrix}
	\xi_{0,0}^{(\upsilon,\vartheta)} & \xi_{0,1}^{(\upsilon,\vartheta)} &  \ldots &  \xi_{0,M-1}^{(\upsilon,\vartheta)}    \\
	\xi_{1,0}^{(\upsilon,\vartheta)} & \xi_{1,1}^{(\upsilon,\vartheta)} & \ldots &  \xi_{1,M-1}^{(\upsilon,\vartheta)}    \\
	\vdots & \vdots & \ddots    &  \vdots       \\
	\xi_{N-1,0}^{(\upsilon,\vartheta)} & \xi_{N-1,1}^{(\upsilon,\vartheta)} & \ldots &   \xi_{N-1,M-1}^{(\upsilon,\vartheta)}
	\end{bmatrix}\right),
	\end{align}
	which consists of all the combinations of frequencies in discrete-time and graph domains as
	\begin{align}\label{JfreqMatrix} 
	\xi_{\ell,k}^{(\upsilon,\vartheta)} = \vartheta \omega_{{\mathsf{G}},\ell} + \upsilon \omega_{\mathsf{D}, k},
	\end{align}   
	where $\ell \in {\llbracket 0,N-1 \rrbracket}, k \in {\llbracket 0,M-1 \rrbracket}$.
	
	\vspace*{0.1cm}
	\textit{Proof: }
	\normalfont{
		From Definition~\ref{JTVtranOper}, one can write
		\begin{align}\label{JTVtransition}
		\mathbf{T}_{\mathsf J}^{(\upsilon,\vartheta)} 
		&= \mathbf{T}_{\mathsf D}^{\upsilon} \otimes \mathbf{T}_{\mathsf G}^{\vartheta} 
		= \left(\mathbf{\Phi}_{\mathsf D} 
		\mathbf{P}_{\mathsf D}^{\upsilon}
		\mathbf{\Phi}_{\mathsf D}^{*}\right) \otimes \left(\mathbf{\Phi}_{\mathsf G} 
		\mathbf{P}_{\mathsf G}^{\vartheta}
		\mathbf{\Phi}_{\mathsf G}^{*}\right) \nonumber\\
		&= \left(\mathbf{\Phi}_{\mathsf D} \otimes \mathbf{\Phi}_{\mathsf G}\right) (\mathbf{P}_{\mathsf D}^{\upsilon} \otimes \mathbf{P}_{\mathsf G}^{\vartheta})
		\left(\mathbf{\Phi}_{\mathsf D} \otimes \mathbf{\Phi}_{\mathsf G}\right)^{*}  \nonumber\\
		&= \mathbf{\Phi}_{\mathsf J}  
		\mathbf{P}_{\mathsf J}^{(\upsilon,\vartheta)} \mathbf{\Phi}_{\mathsf J}^{*},
		\end{align} 
		where the phase shift matrix is defined as
		\begin{align}\label{2020_02_06_09_15}
		\mathbf{P}_{\mathsf J}^{(\upsilon,\vartheta)} =  \mathbf{P}_{\mathsf D}^{\upsilon} \otimes \mathbf{P}_{\mathsf G}^{\vartheta} = \exp\big(-\mi \bm{\mathbf{M}}_{\mathsf J}^{(\upsilon,\vartheta)}\big)
		\end{align} 
		such that
		\begin{align}
		\bm{\mathbf{M}}_{\mathsf J}^{(\upsilon,\vartheta)} = \upsilon \mathbf{M}_{\mathsf D}\oplus \vartheta \mathbf{M}_{\mathsf G}.
		\end{align}
		Using~\eqref{20200624_1838} and~\eqref{20200624_1837}, one can easily write the elements of this matrix as~\eqref{JfreqMatrix}. 
		\hfill $\blacksquare$
	}
\end{thm}

\begin{prop}\label{20200615_0532}
	\normalfont
	Some properties associated with the isometric JTO given by \eqref{JTVtransition} are as follows:
	\begin{enumerate}[(i)]
		\item It is a linear convolutive operator since $\mathbf{P}_{\mathsf J}^{(\upsilon, \vartheta)}$ is a diagonal matrix.
		\item The power spectrum of time-varying signal ${\bm{X}}$ is invariant under the operator $\mathcal{T}_{\mathsf J}^{(\upsilon,\vartheta)}$ as 
		\begin{align}\label{20200921_2039}
		\big|{\widehat{\bm{X}}}^{(\upsilon,\vartheta)}[n,m]\big|^2 = \big|{\widehat{\bm{X}}}[n,m]\big|^2,
		\end{align}
		where ${\widehat{\bm{X}}}[n,m]$ denotes the JFT coefficient of time-varying graph signal $\bm{X}$ corresponding to the frequency pair $(\omega_{\mathsf{G},n-1},\omega_{\mathsf{D},m-1})$ for all $n \in {\llbracket 1,N \rrbracket}, m\in {\llbracket 1,M \rrbracket}$. 
		
		\hspace*{4mm}\textit{Proof:}
		We have
		\begin{align}
		\begin{split}
		\big|{\widehat{\bm{X}}}^{(\upsilon,\vartheta)}[n,m]\big|^2&= \Big|\big(\mathbf{\Phi}_{\mathsf G}^{*}\big(\mathcal{T}_{\mathsf J}^{(\upsilon,\vartheta)} {\bm{X}}\big) {\widebar{\mathbf{\Phi}}_{\mathsf D}}\big) [n,m]\Big|_2^{2} \\
		&= \Big|\big(\mathbf{\Phi}_{\mathsf G}^{*}\mathbf{T}_{\mathsf G}^{\vartheta} {\bm{X}} \big(\mathbf{T}_{\mathsf D}^{\mathsf T}\big)^{\upsilon} {\widebar{\mathbf{\Phi}}_{\mathsf D}}\big) [n,m]\Big|_2^{2} \qquad (\text{cf.}~\eqref{20200921_2033}) \\ 
		&= \big|(\mathbf{P}_{\mathsf G}^{\vartheta}{\widehat{\bm{X}}}\mathbf{P}_{\mathsf D}^{\upsilon}) [n,m] \big|_2^2 \qquad (\text{cf.}~\eqref{DTOFullrepres0907},\eqref{20200617_0828}) 
		\end{split}
		\end{align}
		which clearly reduces to~\eqref{20200921_2039} since $\mathbf{P}_{\mathsf G}^{\vartheta} = \exp(i\vartheta \mathbf{M}_{\mathsf G})$ and $\mathbf{P}_{\mathsf D}^{\upsilon} = \exp(i\upsilon \mathbf{M}_{\mathsf D})$.
		\hfill $\blacksquare$
		
		\item It can be expressed as
		\begin{align}\label{JTOFullrepres0827}
		\mathbf{T}_{\mathsf J}^{(\upsilon,\vartheta)} 
		= \sum_{j=0}^{NM-1} {\upgamma}_{\mathsf{J},j}^{(\upsilon,\vartheta)} {\bm \varphi}_{{\mathsf J}, j} {\bm \varphi}_{{\mathsf J}, j}^{*}, 
		\end{align}
		where for all $j \in {\llbracket 0,NM-1 \rrbracket}$
		\begin{align}\label{JTOAngFreq0733}
		\begin{split}
		{\upgamma}_{\mathsf{J},j}^{(\upsilon,\vartheta)} &= \mathbf{P}_{\mathsf J}^{(\upsilon,\vartheta)}[j+1,j+1] \\ 
		&=\exp\big({-i {\bm{\xi}}^{(\upsilon,\vartheta)}[j+1]}\big) \\
		&= \exp({-i (\vartheta \omega_{{\mathsf{G}},\ell} + \upsilon \omega_{\mathsf{D}, k}})) 
		\end{split}
		\end{align}
		denotes the `$j+1$'-th entry in the main diagonal of $\mathbf{P}_{\mathsf J}^{(\upsilon,\vartheta)}$ (cf.~\eqref{20200617_0929}~\eqref{jointFreq0832},~\eqref{JfreqMatrix}). Note that ${\upgamma}_{\mathsf{J},j}^{(\upsilon,\vartheta)}$ is a bivariate function of $\omega_{{\mathsf{G}},\ell}$ and $\omega_{\mathsf{D}, k}$. 
	\end{enumerate}
\end{prop}

Table~\ref{tab:tbl1} summarizes the characteristics of isometric translation/transition operators in different signal domains including continuous-time, discrete-time, graph, and time-vertex domains where they share similar structural characteristics.

\section{Joint Filtering via Joint Translation Operator}
In discrete-time domain, a linear shift invariant (LSI) filter is equivalent to the circular convolution operator~\cite{vetterli2014Foundations}. Then the filtering operation in this domain can be represented in a compact form as $\mathbf{y} = \mathbf{H}_{\mathsf D} \mathbf{x}$ where $\mathbf{x}$ and $\mathbf{y}$ are the input and output signal vectors, respectively, and the filter matrix can be expressed as follows
\begin{align}
\mathbf{H}_{\mathsf D} 
= \sum_{p=0}^{L_{1}-1} h_{{\mathsf{D}},p} \mathbf{T}_{\mathsf D}^{p},
\end{align}
where $h_{\mathsf{D},0}, \ldots, h_{\mathsf{D},L_1-1}$ are the $L_1$ filter coefficients. In the spectral domain, the dual of $\mathbf{H}_{\mathsf D}$ can be expressed as 
\begin{align}\label{2020_02_06_1120}
\begin{split}
\widehat{\mathbf{H}}_{\mathsf D} 
= \mathbf{\Phi}_{\mathsf D}^{*}\mathbf{H}_{\mathsf D} \mathbf{\Phi}_{\mathsf D}
= \sum_{p=0}^{L_{1}-1} h_{{\mathsf{D}},p} \mathbf{P}_{\mathsf D}^{p}, \quad \text{(cf.~\eqref{DTOFullrepres0907})}.
\end{split}
\end{align}

Filtering in the graph setting is defined as
$
\bm{y} = {\mathbf{H}_\mathsf{G}}\bm{x} = h({\mathbf{L}}_{\mathsf G}) \bm{x}
$
where $\bm{x}$ and $\bm{y}$ are the input and output of the filter, respectively, and the graph filter ${\mathbf{H}_\mathsf{G}} \in \mathbb{C}^{N\times N}$ in the form of matrix function is defined as~\cite{Shuman2013Emerging}
\begin{align}\label{filterkernel}
{\mathbf{H}_\mathsf{G}} = {h}({\mathbf{L}}_{\mathsf G}) = \mathbf{\Phi}_{\mathsf G} {h}({\mathbf{\Lambda}}_{\mathsf G}) \mathbf{\Phi}^{*}_{\mathsf G},
\end{align}
where $h(\mathbf{\Lambda}_{\mathsf G})$ is a diagonal matrix and $h: [0, \lambda_{\mathsf{G},\max}]\to \mathbb{C}$ (cf.~Note~\ref{20200621_1849}). Sandryhaila and Moura~\cite{Sandryhaila2013Discrete,Sandryhaila2014Discrete} considered weighted adjacency matrix $\mathbf{W}_{\mathsf G}$ as the graph shift operator and showed that any LSI graph filter can be written as a polynomial of $\mathbf{W}_{\mathsf G}$~\cite[Theorem~1]{Sandryhaila2013Discrete}. In the same spirit, the graph filter can be written as a polynomial expansion of GTO as follows
\begin{align}\label{graphfilterPoly1212}
\mathbf{H}_{\mathsf G} = \sum_{q=0}^{L_{2}-1} h_{\mathsf{G},q} \mathbf{T}_{\mathsf G}^{q},
\end{align}
where $h_{{\mathsf G},q} \in \mathbb{C}$ is the $q$-th tap of the filter. 
In the graph spectral domain, the dual of $\mathbf{H}_{\mathsf G}$ can be written as 
\begin{align}\label{2020_02_06_1122}
\begin{split}
\widehat{\mathbf{H}}_{\mathsf G} 
= \mathbf{\Phi}_{\mathsf G}^{*}\mathbf{H}_{\mathsf G} \mathbf{\Phi}_{\mathsf G}
= \sum_{q=0}^{L_{2}-1} h_{{\mathsf{G}},q} \mathbf{P}_{\mathsf G}^{q}.
\end{split}
\end{align}

Filtering in the time-vertex domain, or \textit{joint filtering}, is defined as
$
\bm{y} = {\mathbf{H}_\mathsf{J}}\bm{x} = h({\mathbf{L}}_{\mathsf G},{\mathbf{L}}_{\mathsf D}) \bm{x}
$
where $\bm{x}$ and $\bm{y}$ are the input and output of the filter, respectively, and the joint filter is expressed as~\cite{Grassi2018Time,Loukas2016Frequency}
\begin{align}\label{20200621_1701}
{\mathbf{H}_\mathsf{J}} 
= h({\mathbf{L}}_{\mathsf G},{\mathbf{L}}_{\mathsf D}) 
= \mathbf{\Phi}_{\mathsf J} h(\mathbf{\Lambda}_{\mathsf G},\mathbf{\Lambda}_{\mathsf D}) \mathbf{\Phi}_{\mathsf J}^{*},
\end{align}
where $h(\mathbf{\Lambda}_{\mathsf G},\mathbf{\Lambda}_{\mathsf D})$ is a diagonal matrix and $h: [0, \lambda_{\mathsf{G},\max}]\times [0, \lambda_{\mathsf{D},\max}]\to \mathbb{C}$ (cf.~Note~\ref{20200621_1849}). 

\begin{thm}\label{thmjointfilterPoly1807}
	Let $\mathbf{T}_{\mathsf J}^{(\upsilon, \vartheta)}$ be the matrix representation of isometric JTO. A joint filter $\mathbf{H}_{\mathsf J}$ is linear translation invariant (i.e., $\mathbf{T}_{\mathsf J}^{(\upsilon, \vartheta)}\mathbf{H}_{\mathsf J} = \mathbf{H}_{\mathsf J}\mathbf{T}_{\mathsf J}^{(\upsilon, \vartheta)}$)
	if and only if 
	\begin{align}\label{jointfilterPoly1212}
	\mathbf{H}_{\mathsf J} = \sum_{q=0}^{L_{2}-1} \sum_{p=0}^{L_{1}-1} h_{{\mathsf J},(p,q)} \mathbf{T}_{\mathsf J}^{(p,q)},
	\end{align}
	where $h_{{\mathsf J},(p,q)} \in \mathbb{C}$ is the $(p,q)$-th tap of the joint filter and $L_{1}-1$, $L_{2}-1$ are the degrees of bivariate polynomial in discrete-time and graph domains, respectively (i.e., $\mathbf{H}_{\mathsf J}$ is a bivariate polynomial of $\mathbf{T}_{\mathsf J}^{(p,q)}$).
	\vspace{0.1cm}
	
	\textit{Proof:} 
	{\normalfont
		Let us first prove the necessity of this theorem. 
		Let $\mathbf{H}_{\mathsf J}$ be a joint filter which commutes with the bivariate JTO $\mathbf{T}_{\mathsf J}^{(\upsilon, \vartheta)} = \mathbf{\Phi}_{\mathsf J} \mathbf{P}_{\mathsf J}^{(\upsilon, \vartheta)} \mathbf{\Phi}_{\mathsf J}^{*}$ where $\mathbf{P}_{\mathsf J}^{(\upsilon, \vartheta)}$ is a diagonal matrix (cf.~\eqref{JTVtransition},\eqref{2020_02_06_09_15}). Without loss of generality, assume $\upsilon=\vartheta=1$.
		By~\eqref{20200621_1701}, one can write $\mathbf{H}_{\mathsf J} = \mathbf{\Phi}_{\mathsf J} \mathbf{O}_{\mathsf J} \mathbf{\Phi}_{\mathsf J}^{*}$ where $\mathbf{O}_{\mathsf J}$ is a diagonal matrix. Clearly, $\mathbf{H}_{\mathsf J}$ and $\mathbf{T}_{\mathsf J}$ are simultaneously diagonalizable. 
		By~\eqref{JTOAngFreq0733} we have
		\begin{align}
		{\upgamma}_{\mathsf{J},j}= \mathbf{P}_{\mathsf J}[j+1,j+1]  = \exp\big({-i {\bm{\xi}}[j+1]}\big).
		\end{align}
		Suppose $h(x)$, with $x \coloneqq (x_1, x_2)$, is the bivariate polynomial of degrees $L_{1}-1$ and $L_{2}-1$ such that $h({\upgamma}_{\mathsf{J},j}) =\mathbf{O}_{\mathsf J}[j+1,j+1]$ (cf.~\eqref{JTOAngFreq0733}). In other words,
		\begin{align}\label{JFR0623}
		h({\upgamma}_{\mathsf{J},j}) = \sum_{q=0}^{L_{2}-1} \sum_{p=0}^{L_{1}-1} h_{{\mathsf J},(p,q)} {\upgamma}_{\mathsf{J},j}^{(p,q)}, \; \forall j \in {\llbracket 0,NM-1 \rrbracket}, 
		\end{align}
		where $h_{{\mathsf J},(p,q)}$ for all $p,q$ are the polynomial coefficients and ${\upgamma}_{\mathsf{J},j}^{(p,q)}$ is given by~\eqref{JTOAngFreq0733}.
		Therefore,
		\begin{align}
		h(\mathbf{T}_{\mathsf J}) 
		&= \mathbf{\Phi}_{\mathsf J} h(\mathbf{P}_{\mathsf J}) \mathbf{\Phi}_{\mathsf J}^{*} \quad \text{(cf.~Note~\ref{20200621_1849})} \nonumber \\
		&= \sum_{q=0}^{L_{2}-1} \sum_{p=0}^{L_{1}-1} h_{\mathsf{J},(p,q)} \bigg(\sum_{j=0}^{NM-1}{\upgamma}_{\mathsf{J},j}^{(p,q)}{\bm \varphi}_{{\mathsf J}, j} {\bm \varphi}_{{\mathsf J}, j}^{*}\bigg),
		\end{align}
		which clearly reduces to~\eqref{jointfilterPoly1212} by \eqref{JTOFullrepres0827}.
		Next, we prove the sufficiency of the theorem. Since~\eqref{jointfilterPoly1212} holds true, we have 
		\begin{align}
		\mathbf{H}_{\mathsf J} 
		&= \sum_{q=0}^{L_{2}-1} \sum_{p=0}^{L_{1}-1} h_{\mathsf{J},(p,q)} \mathbf{T}_{\mathsf J}^{(p,q)} 
		= \mathbf{\Phi}_{\mathsf J} h({\mathbf{P}_{\mathsf J}}) \mathbf{\Phi}_{\mathsf J}^{*}.
		\end{align}
		Then, one can further deduce that 
		\begin{align}
		\begin{split}
		\mathbf{H}_{\mathsf J} \mathbf{T}_{\mathsf J}
		&= \mathbf{\Phi}_{\mathsf J}\mathbf{O}_{\mathsf J}\mathbf{\Phi}_{\mathsf J}^{*} \mathbf{\Phi}_{\mathsf J} \mathbf{P}_{\mathsf J} \mathbf{\Phi}_{\mathsf J}^{*} \\
		&= \mathbf{\Phi}_{\mathsf J} \mathbf{P}_{\mathsf J}\mathbf{\Phi}_{\mathsf J}^{*} \mathbf{\Phi}_{\mathsf J} \mathbf{O}_{\mathsf J}  \mathbf{\Phi}_{\mathsf J}^{*} 
		= \mathbf{T}_{\mathsf J} \mathbf{H}_{\mathsf J}, 
		\end{split}
		\end{align}
		since $\mathbf{O}_{\mathsf J}$ and $\mathbf{P}_{\mathsf J}$ are diagonal, implying that the joint filter $\mathcal{H}_{\mathsf J}$ is linear translation invariant. 
		\hfill $\blacksquare$ 
	}
\end{thm}

Let ${\boldsymbol{h}}_\mathsf{J} \coloneqq \left[{{h}}_{\mathsf{J},(0,0)},{{h}}_{\mathsf{J},(1,0)}, \ldots, {h}_{\mathsf{J},(L_{1}-1,L_{2}-1)} 
\right]$ be the vector containing the coefficients of joint finite impulse response (JFIR) filter. 
Then it can be written as
\begin{align}\label{20200619_0845}
\mathbf{H}_{\mathsf J} = \mathbf{\Phi}_{\mathsf J} \widehat{\mathbf{H}}_{\mathsf J} \mathbf{\Phi}_{\mathsf J}^{*},
\end{align}
which together with~\eqref{jointfilterPoly1212} leads to its dual in joint spectral domain as 
\begin{align}\label{jointFilterSpectral1450}
\widehat{\mathbf{H}}_{\mathsf J} 
= \sum_{q=0}^{L_{2}-1} \sum_{p=0}^{L_{1}-1} h_{\mathsf{J},(p,q)} \mathbf{P}_{\mathsf J}^{(p,q)}.
\end{align}



\section{Stationarity in Joint Time-Vertex Domain}\label{JTVTopsection}

Let $\mathbf{x}$ be a TWSS process in discrete-time domain and ${\mathbf R}_{\mathbf{x}} = \mathbb{E}[\mathbf{x}\mathbf{x}^{*}]$ be its autocorrelation matrix. It is known that isometric translation operator in discrete-time domain $\mathbf{T}_{\mathsf D}$ and autocorrelation matrix ${\mathbf R}_{\mathbf{x}}$ are simultaneously diagonalizable by the DFT matrix.

\begin{defn}(Multivariate TWSS~\cite{Bach2004Learning})\label{MTWSS}
	Suppose that ${\mathbf{x}}_{m}\in{\mathbb R}^N$ is a vector random process in discrete-time $m$. 
	Let $\mathbf{X} = [{\mathbf{x}}_{1}, {\mathbf{x}}_{2}, \ldots,{\mathbf{x}}_{M}] \in \mathbb{R}^{N \times M}$ be the collection of such random vectors. It is called 
	multivariate TWSS (MTWSS) if and only if 
	\begin{enumerate}[(i)]
		\item $\mathbb{E}[\mathbf{x}_m] = c\mathbf{1}_{N}$ for all $m$; 
		\item The autocorrelation matrix for every two time indices $m$ and $n$ is  ${\bm{\mathsf{R}}}_{\mathsf{D},\left(m,n\right)} = \mathbb{E}[\mathbf{x}_{m}\mathbf{x}_{n}^*] = \mathbf{\Psi}_{(m-n) \; {\mathsf{mod}} \; M}$.
	\end{enumerate}
\end{defn}


\begin{remk}\label{20200601_1316}
	Here, we provide an alternative representation of Definition~\ref{MTWSS}. Suppose that ${\mathbf{x}}_{m}\in{\mathbb R}^N$ is a vector random process in $m$. 
	Let $\mathbf{X} = [{\mathbf{x}}_{1}, {\mathbf{x}}_{2}, \ldots,{\mathbf{x}}_{M}] \in \mathbb{R}^{N \times M}$ be the collection of such random vectors. It is called 
	MTWSS if and only if 
	\begin{enumerate}[(i)]
		\item $\mathbb{E}[\mathbf{X}] = \mathbb{E}[\mathbf{X}(\mathbf{T}_{\mathsf D}^\mathsf{T})^{\upsilon}]$; 
		\item ${\mathbf R}_{{\mathbf{X}}} = \left[{\bm{\mathsf{R}}}_{\mathsf{D},\left(m,n\right)}\right]$ 
		such that ${\bm{\mathsf{R}}}_{\mathsf{D},\left(m,n\right)} = \mathbb{E}[\mathbf{x}_{m}\mathbf{x}_{n}^*] =  \mathbf{\Psi}_{(m-n) \; {\mathsf{mod}} \; M}$.
		\hfill $\openbox$ 
	\end{enumerate}
\end{remk}

\begin{defn}(\cite[Definition 16]{girault2015Signal})\label{GiraultStationary}
	Let $\bm{x}$ be a stochastic graph signal on graph $\mathsf G$ where $\bm{x}[n]$ is the random variable corresponding to the vertex $n$. Then $\bm{x}$ is called vertex WSS (VWSS) with respect to (w.r.t.) the translation operator on graph $\mathbf{T}_{\mathsf G}^{\vartheta}$ if and only if for all $\vartheta$:
	\begin{enumerate}[(i)]
		\item ${\mathbf m}_{\bm{x}} = \mathbb{E}[\bm{x}] = \mathbb{E}[\mathbf{T}_{\mathsf G}^{\vartheta} \bm{x}]$;
		\item ${\mathbf R}_{\bm{x}} = \mathbb{E}[\bm{x}\bm{x}^{*}] = \mathbb{E}\big[(\mathbf{T}_{\mathsf G}^{\vartheta} \bm{x})\big(\mathbf{T}_{\mathsf G}^{\vartheta} \bm{x}\big)^{*}\big].$
	\end{enumerate}
\end{defn}

Via different approaches for defining wide-sense stationarity on graph, it is shown~\cite{Girault2015Stationary} or assumed~\cite{Perraudin2017Stationary,Marques2017Stationary} that the GFT matrix diagonalizes the autocorrelation matrix as ${\mathbf R}_{\bm{x}} = \mathbf{\Phi}_{\mathsf G} \mathbf{S}_{\bm x} \mathbf{\Phi}_{\mathsf G}^{*}$ where ${\mathbf S}_{\bm x}$ is a diagonal GPSD matrix consistent with the time domain.

\begin{defn}(Multivariate VWSS)\label{MVWSS}
	Let ${\bm{x}}_{m}\in{\mathbb R}^N$ be a vector random process in $m$ and $\bm{X} = [{\bm{x}}_{1}, {\bm{x}}_{2}, \ldots,{\bm{x}}_{M}] \in \mathbb{R}^{N \times M}$ be the collection of such random vectors. It is called 
	multivariate VWSS (MVWSS) if and only if 
	\begin{enumerate}[(i)] 
		\item $\mathbb{E}[\bm{X}] = \mathbb{E}[\mathbf{T}_{\mathsf G}^{\vartheta} \bm{X}]$; 
		\item ${\mathbf R}_{{\bm{X}}} =  \left[{\bm{\mathsf{R}}}_{\mathsf{G},\left(m,n\right)}\right]$ where ${\bm{\mathsf{R}}}_{\mathsf{G},\left(m,n\right)} = \mathbb{E}[\bm{x}_{m}\bm{x}_{n}] = \mathbb{E}\big[(\mathbf{T}_{\mathsf G}^{\vartheta} \bm{x}_m)\big(\mathbf{T}_{\mathsf G}^{\vartheta} \bm{x}_n\big)^{*}\big]$ such that 
		\begin{align}
		{\bm{\mathsf{R}}}_{\mathsf{G},\left(m,n\right)} = \mathbf{\Phi}_{\mathsf G} \widehat{\bm{\mathsf{R}}}_{\mathsf{G},\left(m,n\right)} \mathbf{\Phi}_{\mathsf G}^{*}, \quad m,n \in {\llbracket 1,M \rrbracket},
		\end{align}
		and $\widehat{\bm{\mathsf{R}}}_{\mathsf{G},\left(m,n\right)}$ is a diagonal matrix. 
	\end{enumerate}
\end{defn}

The bivariate JTO (Definition~\ref{JTVtranOper}) enables us to generalize wide-sense stationarity to the time-vertex domain as follows. 

\begin{defn}\textit{(JWSS via Bivariate translation Invariance)}\label{JWSSdef2}
	A zero-mean stochastic process $\bm{x} = {\mathsf{vec}}(\bm{X})$ defined on graph $\mathsf{G}$,  where $\bm{X} \in \mathbb{R}^{N \times M}$, is called JWSS under $\mathbf{T}_{\mathsf J}^{(\upsilon,\vartheta)}$ (cf.~Definition~\ref{JTVtranOper}) if and only if for all $\vartheta$ and $\upsilon$ we have 
	\begin{align}
		{\mathbf R}_{\bm{x}} = \mathbb{E}[\bm{x}\bm{x}^{*}] = \mathbb{E}\big[\big(\mathbf{T}_{\mathsf J}^{(\upsilon,\vartheta)}  \bm{x}\big)\big(\mathbf{T}_{\mathsf J}^{(\upsilon,\vartheta)} \bm{x}\big)^{*}\big].
	\end{align}
\end{defn}

\begin{thm}\label{20200622_1944}
	Let $\bm{x}$ be a zero-mean stochastic process on a connected graph $\mathsf{G}$. Then $\bm{x}$ is JWSS based on the isometric JTO $\mathcal{T}_{\mathsf J}^{(\upsilon,\vartheta)}$, according to Definition~\ref{JWSSdef2}, if and only if the autocorrelation matrix is unitarily diagonalizable by the JFT matrix as 
	\begin{align}\label{20200622_2028}
	{\mathbf R}_{{\bm{x}}} = {\mathbf{\Phi}_{\mathsf J}} {\mathbf S}_{\bm x} \mathbf{\Phi}_{\mathsf J}^{*},
	\end{align} 
	where ${\mathbf S}_{\bm x}$ is a diagonal matrix with nonnegative real entries on its main diagonal{\footnote{~The proof is inspired by: ~{\bf\small{\texttt{\url{ https://dsp.stackexchange.com/a/68284/17094/}}}})}}.
	\vspace*{0.1cm}
	
	\textit{Proof}
	{\normalfont 
		By Definition~\ref{JWSSdef2}~(ii), the autocorrelation matrix is written as
		\begin{equation}\label{JWSSspactralautocorr832}
		\begin{split}
		{\mathbf R}_{\bm{x}} = \mathbb{E}[{\bm{x}}{\bm{x}}^{*}] = \mathbb{E}\big[\big({\mathbf{T}_{\mathsf J}^{(\upsilon,\vartheta)}\bm{x}}\big) \big(\mathbf{T}_{\mathsf J}^{(\upsilon,\vartheta)}{\bm{x}}\big)^{*}\big] 
		= \mathbf{T}_{\mathsf J}^{(\upsilon,\vartheta)} \mathbb{E}[{\bm{x}}{\bm{x}}^{*}] \mathbf{T}_{\mathsf J}^{{(-\upsilon,-\vartheta)}}.
		\end{split}
		\end{equation}
		Since the bivariate JTO is unitary (cf.~Proposition~\ref{20200622_1443}), one can readily see that
		$
		{\mathbf R}_{\bm{x}} \mathbf{T}_{\mathsf J}^{(\upsilon,\vartheta)} = \mathbf{T}_{\mathsf J}^{(\upsilon,\vartheta)} {\mathbf R}_{\bm{x}},
		$
		meaning the autocorrelation matrix and bivariate JTO commute. Moreover, we know that the autocorrelation matrix is a positive semidefinite matrix (in other words, diagonalizable with non-negative eigenvalues). Then, $\mathbf{T}_{\mathsf J}^{(\upsilon,\vartheta)}$ and ${\mathbf R}_{\bm{x}}$ are diagonalizable matrices and commute, implying that they are simultaneously diagonalizable~\cite[Theorem~1.3.12]{Horn1985Matrix}. Since the bivariate JTO $\mathbf{T}_{\mathsf J}^{(\upsilon,\vartheta)}$ is diagonalizable by the JFT matrix (cf.~\eqref{JTVtransAbstract}), the result~\eqref{20200622_2028} follows immediately. For the proof of sufficiency, by the reverse implications above, it can be easily shown that $\bm{x}$ is JWSS based on the bivariate JTO. 
	}
	\hfill $\blacksquare$
\end{thm}

\begin{remk}(See also~\cite[Property~2]{Loukas2017JStationary}))\label{JFJWSS}
	Let $\mathbf{H}_{\mathsf J}$ be a linear JFIR filter defined over the connected graph $\mathsf{G}$ and ${\bm{x}}$ be the zero-mean JWSS process under the JTO $\mathcal{T}_{\mathsf J}^{(\upsilon,\vartheta)}$. Then the output of this filter is a JWSS process under the JTO as well. 
	
	\vspace*{0.1cm}
	\textit{Proof:}
	{\normalfont  
	Let the output of joint filter $\mathbf{H}_{\mathsf J}$ be ${\bm{y}} = \mathbf{H}_{\mathsf J} {\bm{x}}$. Then we have
	\begin{align}\label{AutoCorrMatJWSSJF}
	\begin{split}
	{\mathbf R}_{\bm{y}} 
	&= \mathbb{E}[(\mathbf{H}_{\mathsf J} \bm{x})(\mathbf{H}_{\mathsf J} \bm{x})^{*}] 
	= \mathbf{H}_{\mathsf J} \mathbb{E}[\bm{x}\bm{x}^{*}] \mathbf{H}_{\mathsf J}^{*}  \\
	&= {\mathbf{\Phi}_{\mathsf J}} \widehat{\mathbf{H}}_{\mathsf J}\mathbf{\Phi}_{\mathsf J}^{*} {\mathbf R}_{\bm{x}} {\mathbf{\Phi}_{\mathsf J}} \widehat{\mathbf{H}}_{\mathsf J}^{*}\mathbf{\Phi}_{\mathsf J}^{*} 
	= {\mathbf{\Phi}_{\mathsf J}} \widehat{\mathbf{H}}_{\mathsf J} {\mathbf S}_{{\bm{x}}} \widehat{\mathbf{H}}_{\mathsf J}^{*}\mathbf{\Phi}_{\mathsf J}^{*} \\
	&= {\mathbf{\Phi}_{\mathsf J}} \widehat{\mathbf{H}}_{\mathsf J} \widehat{\mathbf{H}}_{\mathsf J}^{*} {\mathbf S}_{\bm{x}} \mathbf{\Phi}_{\mathsf J}^{*} 
	= {\mathbf{\Phi}_{\mathsf J}} {\mathbf S}_{{\bm{y}}} \mathbf{\Phi}_{\mathsf J}^{*},
	\end{split}
	\end{align}
	since $\widehat{\mathbf{H}}_{\mathsf J}$ and ${\mathbf S}_{\bm{x}}$ are diagonal (cf.~\eqref{20200622_2028}), where
	\begin{align}\label{20200207_0830}
	{\mathbf S}_{\bm{y}} = |\widehat{\mathbf{H}}_{\mathsf J}|^{2} {\mathbf S}_{{\bm{x}}},
	\end{align}
	is the key equation relating the input JPSD to the output JPSD of the joint filter. 
	\hfill $\blacksquare$ 
	}
\end{remk}

An important question concerning JWSS processes may arise here: ``\textit{What is the relation between JWSS and classical multivaruiate WSS (MWSS) processes in time and vertex domains?}" Loukas and Perraudin~\cite[Theorem~2]{Loukas2017JStationary} showed that a JWSS process, defined via joint filtering, is both MTWSS and MVWSS. In the following theorem, we elaborate on the relation between JWSS and classical MWSS processes via translation invariance. 

\begin{thm}\label{MTWSSMVWSS}
	A zero-mean time-vertex process $\bm{x}$ over a connected graph $\mathsf{G}$ is JWSS under the JTO $\mathcal{T}_{\mathsf J}^{(\upsilon,\vartheta)}$ if and only if it is simultaneously MTWSS and MVWSS based upon~Definition~\ref{MTWSS} and Definition~\ref{MVWSS}. \vspace{0.1cm}
	
	\textit{Proof:}
	{\normalfont 
	The necessity of this theorem for the first-moment is clearly true.
	For the second-order moment, by~\eqref{20200622_2028}, we have ${\mathbf R}_{{\bm{x}}} = {\mathbf{\Phi}_{\mathsf J}} \mathbf{S}_{\bm x} \mathbf{\Phi}_{\mathsf J}^{*}$, where the JPSD matrix can be written as
	\begin{align}
	\mathbf{S}_{\bm x} = \widehat\oplus_{j=1}^n {\bm{\mathsf{S}}}_{j}, 
	\end{align}
	which is the direct sum of diagonal matrices ${\bm{\mathsf{S}}}_{j}$ with nonnegative entries on the main diagonal. On the other hand, by~\eqref{JFTMatrix1655}, ${\mathbf{\Phi}_{\mathsf J}}$ can be described as a block matrix 
	${\mathbf{\Phi}_{\mathsf J}} = [{\bm{\mathsf{\Phi}}}_{(m,n)}]$ where 
	\begin{align*}
	{\bm{\mathsf{\Phi}}}_{(m,n)} = \exp(\mi 2\pi (m-1)(n-1)/{M}) {\mathbf{\Phi}_{\mathsf G}} \in \mathbb{C}^{N \times N}
	\end{align*}
	is its submatrix in $m$-th row and $n$-th column partition, and $m,n \in {\llbracket 1,M \rrbracket}$. Then ${\mathbf R}_{{\bm{x}}}$ can be re-expressed as the following block matrix  
	\begin{align}
	{\mathbf R}_{{\bm{x}}} = 
	\begin{bmatrix}
	\mathbf{\Xi}_{(1,1)} &  \mathbf{\Xi}_{(1,2)} & \ldots &   \mathbf{\Xi}_{(1,M)} \\
	\mathbf{\Xi}_{(2,1)} &  \mathbf{\Xi}_{(2,2)} & \ldots &   \mathbf{\Xi}_{(2,M)} \\
	\vdots               &  \vdots & \ddots      &  \vdots    \\
	\mathbf{\Xi}_{(M,1)} &  \mathbf{\Xi}_{(M,2)} & \ldots &   \mathbf{\Xi}_{(M,M)}
	\end{bmatrix},
	\end{align}
	where 
	\begin{align}\label{blockDiagAutoCorr}
	\mathbf{\Xi}_{(m,n)} = \mathbf{\Phi}_{\mathsf G} \widehat{\mathbf{\Xi}}_{(m,n)} \mathbf{\Phi}_{\mathsf G}^{*}, \quad m,n \in {\llbracket 1,M \rrbracket},
	\end{align}
	and
	\begin{align}
	\widehat{\mathbf{\Xi}}_{(m,n)} = \frac{1}{M}\sum_{j=1}^{M} {\bm{\mathsf{S}}}_{j} \exp\left(\frac{\mi 2\pi(m-n)}{M}\right),
	\end{align}
	implying that $\mathbf{\Xi}_{(m,n)}$ depends on the discrete time difference $m-n$. Clearly, ${\mathbf R}_{{\bm{x}}}$ is a \textit{block circulant matrix} and hence $\bm{x}$ is a MTWSS process (cf. (ii) in~Remark~\ref{20200601_1316}). Furthermore, by~\eqref{blockDiagAutoCorr}, the submatrices of ${\mathbf R}_{{\bm{x}}}$, namely $\mathbf{\Xi}_{(m,n)}$ for all $n,m$ are simultaneously diagonalizable with the graph Laplacian. Therefore, it is MVWSS (cf. (ii) in Definition~\ref{MVWSS}). On the other hand, for the proof of sufficiency, assuming that the process $\bm{x}$ is simultaneously MTWSS and MVWSS, by the reverse implications above, it can be easily shown that $\bm{x}$ is JWSS. 	
	\hfill $\blacksquare$ 
	}
\end{thm}

\begin{disc}\label{20200616_0532}
	Let $\bm{x}$ be a zero-mean JWSS process under the JTO $\mathcal{T}_{\mathsf J}^{(\upsilon,\vartheta)}$. Then, $\bm{x}$ can be written as the output of a joint time-vertex filter $\mathbf{H}_{\mathsf J}$ (cf.~\eqref{20200619_0845}) in response to a white noise $\bm{z}$ with zero-mean and autocorrelation matrix $\mathbf{R}_{\bm{z}} = \mathbf{I}_{NM}$. One can deduce that $\mathbf{S}_{\bm{x}} = |\widehat{\mathbf{H}}_{\mathsf J}|^{2}$ (cf.~\eqref{20200207_0830}). For the simplicity of notation, let $\theta_j\coloneqq\mathbf{S}_{\bm{x}}[j+1,j+1]$ be the $j$-th JPSD component where $j \in {\llbracket 0,NM-1 \rrbracket}$. Then by~\eqref{jointFilterSpectral1450}, we can write
	\begin{align}\label{20200617_1123}
	\theta_j = \Big|\sum_{q=0}^{L_{2}-1} \sum_{p=0}^{L_{1}-1} h_{\mathsf{J},(p,q)} {\upgamma}_{\mathsf{J},j}^{(p,q)}\Big|^2 = |h(\omega_{{\mathsf{G}},\ell}, \omega_{\mathsf{D}, k})|^2,
	\end{align}
	where $h: (\omega_{{\mathsf{G}},\ell},\omega_{\mathsf{D}, k}) \to \mathbb{C}$ is a bivariate function and ${\upgamma}_{\mathsf{J},j}^{(p,q)} = \exp({-i (q \omega_{{\mathsf{G}},\ell} + p \omega_{\mathsf{D}, k}}))$ (cf.~\eqref{JTOAngFreq0733},~\eqref{20200617_0929}). It implies that $\theta_j$ for all $j$ is a bivariate function of $\omega_{{\mathsf{G}},\ell}$ and $\omega_{\mathsf{D}, k}$. On the other hand, if one chooses the isometric GTO --- using joint Laplacian matrix, as the translation (or translation) operator on joint graph (cf.~\eqref{20200616_1658},~\eqref{20200617_0929}) --- then the $j$-th component of JPSD obtained via this approach is
	\begin{align}\label{20200617_1124}
	\theta_j^{\prime} = |g(\lambda_{{\mathsf J}, j})|^{2}, 
	\end{align}
	where $g: \lambda_{{\mathsf J}, j} \to \mathbb{C}$. Clearly, by defining the stationarity on joint graph using isometric GTO, the JPSD is a univariate function of eigenvalues of joint Laplacian matrix, i.e., $\lambda_{{\mathsf J}, j}$ (cf.~\eqref{20200616_1658}). From~\eqref{20200617_1123},~\eqref{20200617_1124}, and Theorem~\ref{20200622_1944} one can infer that the JWSS process, in the second-order sense, is more general than wide-sense stationarity on joint graph where the latter is actually a special case of former.  
	The reader is referred to~\cite{Loukas2017JStationary} for a further detailed discussion about the difference between the general notion of stationarity in time-vertex domain and stationarity on joint graph.
	\hfill $\openbox$
\end{disc}

\section{Joint Power Spectral Density Estimation}
\label{section:JPSD_Estimators}
Analogous to the stochastic processes in Euclidean space, a reliable JPSD estimator is important for analyzing time-series on graph. In this section, given a data $\mathscr{X} \coloneqq \{\bm{x}_{q}, \; q\in {\llbracket 1,Q \rrbracket}\}$ where $Q$ is the number of realizations, we present the estimation of JPSD vector denoted by 
\begin{align}\label{20200624_1742}
{\boldsymbol{\uptheta}}_{\bm x} \coloneqq {\mathsf{diag}}\left(\mathbf{S}_{\bm x}\right)
\end{align}
of a JWSS process ${\bm x} = {\mathsf{vec}}(\bm{X})$. The generalized Welch method (GWM) is defined as the average of windowed periodograms (cf.~\eqref{GWMJPSD1837} below). 
So, we begin with the definition of windowing in joint time-vertex domain.

\begin{defn}(\textit{Joint Windowing})\label{20200207_1343}
	Let ${\bm{x}} = {\mathsf{vec}}\left({\bm{X}}\right)$ be the given time-varying graph signal. Let $\bm{A}_{\mathsf D} = {\mathsf{Diag}}(\bm{a}_{\mathsf D})$ and $\bm{A}_{\mathsf G} = {\mathsf{Diag}}(\bm{a}_{\mathsf G})$ be the windowing matrices corresponding to the windows $\bm{a}_{\mathsf D}$ and $\bm{a}_{\mathsf G}$ in discrete-time and graph domains, respectively. The time-vertex windowing is defined as
	$	
	{{\bm{X}}_{\mathsf w}} \coloneqq \bm{A}_{\mathsf G} {\bm{X}} \left(\bm{A}_{\mathsf D}^{\mathsf T}\right).
	$
	In the vector form, ${\bm{x}_{\mathsf{w}}} \coloneqq \bm{a}_{\mathsf J}  \odot {\bm{x}} = \bm{A}_{\mathsf J}{\bm{x}}$ where
	\begin{align}\label{JWmatrix1231}
	\bm{A}_{\mathsf J} = {\mathsf{Diag}}(\bm{a}_{\mathsf J}) = \bm{A}_{\mathsf D} \otimes \bm{A}_{\mathsf G},
	\end{align}
	is the joint window matrix and ${\bm{x}_{\mathsf{w}}} = {\mathsf{vec}}({\bm{X}}_{\mathsf{w}})$. 
\end{defn}

By this definition, we have $\widehat{{\bm{x}}}_{\mathsf{w}} = \widehat{\bm{A}}_{\mathsf J} \widehat{\bm{x}}$ where
\begin{align}\label{20200207_1332}
\widehat{\bm{A}}_{\mathsf J} = \mathbf{\Phi}_{\mathsf J}^{*} \bm{A}_{\mathsf J} \mathbf{\Phi}_{\mathsf J} = \mathbf{\Phi}_{\mathsf J}^{*} {\mathsf{Diag}}(\bm{a}_{\mathsf J}) \mathbf{\Phi}_{\mathsf J}
\end{align}
is the dual joint windowing matrix in spectral domain. 

Following this definition, the generalized Welch JPSD estimator ${\accentset{\circ}{\mathbf{\uptheta}}}_{{\bm x},\mathsf{GWM}}$ is defined as:
\begin{align}\label{GWMJPSD1837} 
{\accentset{\circ}{\mathbf{\uptheta}}}_{{\bm x},\mathsf{GWM}} [k] \coloneqq \frac{1}{Q} \sum_{q=1}^{Q} \left|(\mathbf{\Phi}_{\mathsf J}^{*} \bm{A}_{{\mathsf J}} \bm{x}_{q}) [k] \right|^{2}.
\end{align}
It is worth noting that by assuming $\bm{A}_{{\mathsf J}} = \mathbf{I}_{NM}$ in~\eqref{GWMJPSD1837}, GWM reduces to the generalized Bartlett method (GBM) --- also called sample estimator~\cite{Loukas2017JStationary}, denoted by ${\accentset{\circ}{\mathbf{\uptheta}}}_{{\bm x},\mathsf{GBM}}$ in Section~\ref{Experimental}, as an unbiased estimator. 
The next theorem provides the bias and variance of generalized Welch JPSD estimator. 

\begin{remk}\label{20200208_1028}
	For a given single realization of a JWSS process denoted by $\bm{x} = \mathsf{vec}(\bm{X}) \in \mathbb{C}^{NM}$, we exploit a bank of joint windows for JPSD estimation.
	In discrete-time domain, the time-series data of length $M$ is split up into overlapping segments of length $L$ where 
	\begin{align}\label{20200208_1021}
	K_1 \coloneqq \left\lfloor \frac{M- L}{\Delta\tau} \right\rfloor + 1,
	\end{align}
	is the number of windows, $\lfloor\cdot\rfloor$ is the floor function, and $\Delta\tau$ is the length of overlap.  
	By this, we have a set of discrete-time windows ${\mathscr{A}}_{\mathsf D} \coloneqq \{\bm{A}_{{\mathsf D}, k_1} : k_1 \in {\llbracket 1, K_1 \rrbracket}\}$. 
	Moreover, following the same concept of local windowing~\cite{Marques2017Stationary}, we obtain a set of graph windows 
	${\mathscr{A}}_{\mathsf G} \coloneqq \{\bm{A}_{{\mathsf G}, k_2} : k_2 \in {\llbracket 1, K_2 \rrbracket}\}$. Then we come up with a bank of joint windows as follows
	\begin{align}\label{20200208_1022}
	{\mathscr{A}}_{\mathsf J} \coloneqq \{\bm{A}_{{\mathsf J}, k} = \bm{A}_{{\mathsf D}, k_1}\otimes\bm{A}_{{\mathsf G}, k_2}: k \in {\llbracket 1, K_1 K_2 \rrbracket}\}.
	\end{align}
	Then, we calculate
	\begin{align}\label{20200208_0914} 
	{\accentset{\circ}{\mathbf{\uptheta}}}_{{\bm x},\mathsf{GWM}}[\ell] = \frac{1}{K_1 K_2} \sum_{k=1}^{K_1 K_2} \big|(\mathbf{\Phi^*_{\mathsf G}}{\bm{A}}_{{\mathsf J},k} {\bm{x}}) [\ell]\big|^{2},
	\end{align}
	where $\ell \in {\llbracket 1, NM \rrbracket}$. 
	\hfill $\openbox$
\end{remk}

\section{Simulation and Experimental Results}
\label{Experimental}
This section presents simulation and experimental results to demonstrate the effectiveness of the proposed JPSD estimator ${\accentset{\circ}{\mathbf{\uptheta}}}_{{\bm x},\mathsf{GWM}}$ given by~\eqref{GWMJPSD1837} and usefulness of JWSS modeling. 

\raggedright
\justify
{${\mathsf{\mathbf{Simulation~Results.}}}$}
We generate the time-series on graph, of length $M=128$, by passing the white Gaussian noise through a chosen joint filter of degrees $L_1-1$ and $L_2-1$ in discrete-time and graph domains, respectively (cf.~\eqref{jointfilterPoly1212},~\eqref{20200619_0845}, and~\eqref{jointFilterSpectral1450}). Likewise~\cite{Marques2017Stationary}, we also use Erdős–Rényi  graph~\cite{bollobas2001Random} and Watts-Strogatz small-world graph~\cite{Watts1998Collective} with $N \in \{100,200\}$ nodes in our simulations. In discrete-time window, we use the Hamming window with $50\%$ overlapping (cf.~\eqref{20200208_1021},~\eqref{20200208_1022}). Then, with the obtained bank of joint windows stated in Remark~\ref{20200208_1028}, we calculate ${\accentset{\circ}{\mathbf{\uptheta}}}_{{\bm x},\mathsf{GWM}}$ (cf.~\eqref{20200208_0914}). Then, for the estimated JPSD ${\accentset{\circ}{\mathbf{\uptheta}}}_{{\bm x}}$ (via GBM or GWM), we compute the normalized mean-squared error (NMSE), bias, and standard deviation (Std) as follows:
$$
\mathsf{NMSE} = {\widetilde{\mathbb{E}}\big[\|{\accentset{\circ}{\boldsymbol{\uptheta}}}_{{\bm x}} - \boldsymbol{\uptheta}_{\bm x}\|_{2}^2\big]}/{\|\boldsymbol{\uptheta}_{\bm x}\|_{2}^2},
$$
$$
\mathsf{Bias} = {\big\|\widetilde{\mathbb{E}}\big[{\accentset{\circ}{\boldsymbol{\uptheta}}}_{{\bm x}}\big] - \boldsymbol{\uptheta}_{\bm x}\big\|_{2}}/{\|\boldsymbol{\uptheta}_{\bm x}\|_{2}},
$$
$$
\mathsf{Std} = {\Big(\widetilde{\mathbb{E}}\Big[\big\|{\accentset{\circ}{\boldsymbol{\uptheta}}}_{{\bm x}} - \widetilde{\mathbb{E}}\big[{\accentset{\circ}{\boldsymbol{\uptheta}}}_{{\bm x}}\big]\big\|_{2}^2\Big]\Big)^{1/2}}/{\|\boldsymbol{\uptheta}_{\bm x}\|_{2}},
$$
where $\widetilde{\mathbb{E}}[\cdot]$ is the average over all the realizations.


Figure~\ref{fig:NMSE_GBM_ER_WS} depicts the NMSE performance of JPSD estimator ${\accentset{\circ}{\mathbf{\uptheta}}}_{{\bm x},\mathsf{GBM}}$ of $Q \in {\llbracket 1,10 \rrbracket}$ independent realizations of JWSS process. The Erdős–Rényi and Watts-Strogatz graphs of $N \in \{100,200\}$ nodes are used, and meanwhile the true JPSD is obtained from a JFIR filter of degrees $L_1 = 7$ and $L_2 = 4$. For all the considered cases shown in this figure, NMSE performance of GBM is better for larger $Q$. However, this estimator suffers from high NMSE when only a small number of realizations of the process are available. Next, we will focus on the case where there exists only a single realization of the process. 

\begin{figure}[t!]
	\centering
	\vspace*{-4mm} 
	\hspace*{-6mm}
	\includegraphics[scale=0.85]{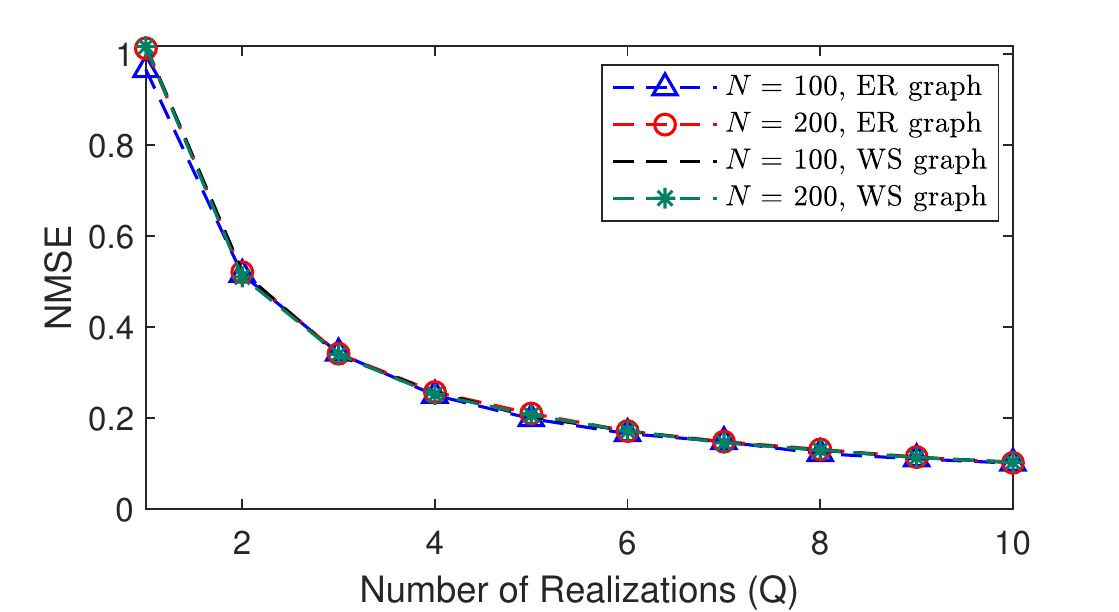}
	\caption[fig1]{NMSE performance of JPSD estimator ${\accentset{\circ}{\mathbf{\uptheta}}}_{{\bm x},\mathsf{GBM}}$ for $L_1 = 7$ and $L_2 = 4$ where Watts-Strogatz and Erdős–Rényi graphs with $N \in \{100,200\}$ vertices are used.}
	\label{fig:NMSE_GBM_ER_WS}
\end{figure}

Figure~\ref{fig:NMSE_vs_L1L2_SWG_N100} presents the NMSE performance of ${\accentset{\circ}{\mathbf{\uptheta}}}_{{\bm x},\mathsf{GWM}}$ versus the degree pairs of JFIR filters $(L_1,L_2)$. Here, there is only a single realization of JWSS process available. The Erdős–Rényi and Watts-Strogatz graphs with $N \in \{100,200\}$ vertices are used. Moreover, the simulation is performed for window number pairs $(K_1, K_2) \in \{(7,5),(7,10)\}$. It can be seen that, for all scenarios, GWM performs significantly better than GBM for $Q=1$ as shown in Figure~\ref{fig:NMSE_GBM_ER_WS}. The trend of NMSE linearly and mildly increases with $L_1$ and $L_2$ for all the considered cases indicating its efficacy for both graphs.

Figure~\ref{fig:20200216_1252} illustrates the performance of GWM w.r.t. the length of discrete-time window $L$ and number of windows in graph setting $K_2$ (cf.~\eqref{20200208_1021},~\eqref{20200208_0914}). In this simulation, there is only a single realization of JWSS process and the Erdős–Rényi graph with $N = 200$ nodes is used. Overall, from Figure~\ref{20200217_0534},~\ref{20200217_0535} and~\ref{20200217_0536}, one can observe that the larger the $L$, the better the NMSE performance along with lower bias and higher Std, whereas the variation of NMSE versus $K_2$ is mild indicating its low sensitivity to $K_2$. Furthermore, it can be observed from Figure~\ref{fig:20200216_1252}(d), that it is also computationally efficient for smaller $K_2$ and larger $L$.

\raggedright
\justify
{${\mathsf{\mathbf{Experimental~Results~(ER).}}}$}
To show the practicality of JWSS modeling,~\cite{Loukas2017JStationary} focused on the recovery of JWSS processes while~\cite{Isufi2019Forecasting} concentrated on forecasting of time-series on graphs. In our experiment, we present the usefulness of JWSS modeling for the classification of ``time-series on graph" via JPSD features, using real EEG data, which can potentially be a new application for graph signal processing. To this end, we model EEG signals as JWSS processes for the purpose of: (1) Emotion recognition and (2) Alzheimer’s disease (AD) recognition from EEG data. To the best of our knowledge, Dong~\cite{Dong2014Multi} is the first who addressed the classification of ``graph signals", using features derived from graph spectral information, where the principal component analysis (PCA) and support vector machine (SVM) are used for feature reduction and classification, respectively. Likewise, it is sufficient to exploit the PCA and SVM as the fundamental machine learning tools (rather than advanced tools) in our experiment as in~\cite{Dong2014Multi}.

\begin{figure}[t!]
	\centering
	\vspace*{-7mm}
	\hspace*{-3mm}
	\subfloat[]{
		\includegraphics[scale=0.67]{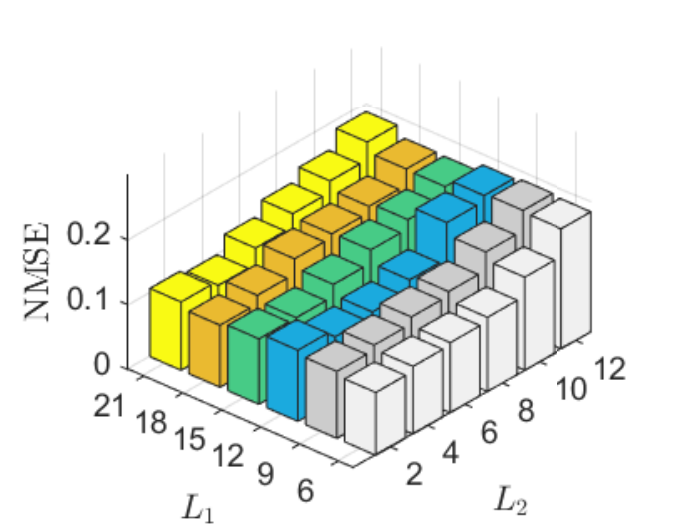}
		\label{20200212_1608}
	}
	\hspace*{-8mm}
	\subfloat[]{
		\includegraphics[scale=0.74]{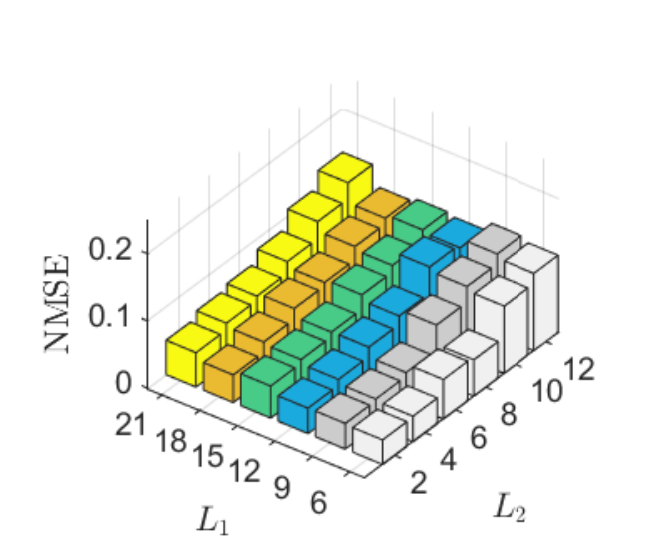}
		\label{20200212_1609}
	} 
	\hspace*{-4mm}
	\subfloat[]{
		\includegraphics[scale=0.60]{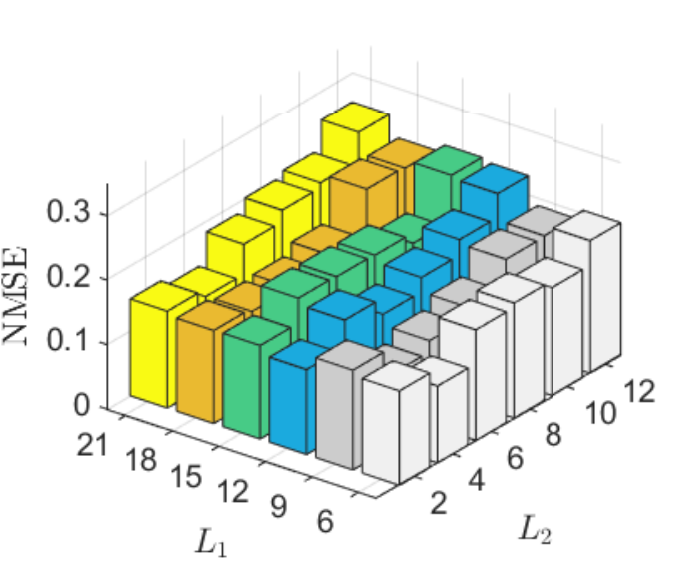}
		\label{20200625_0559}
	}
	\hspace*{-5mm}
	\subfloat[]{
		\includegraphics[scale=0.65]{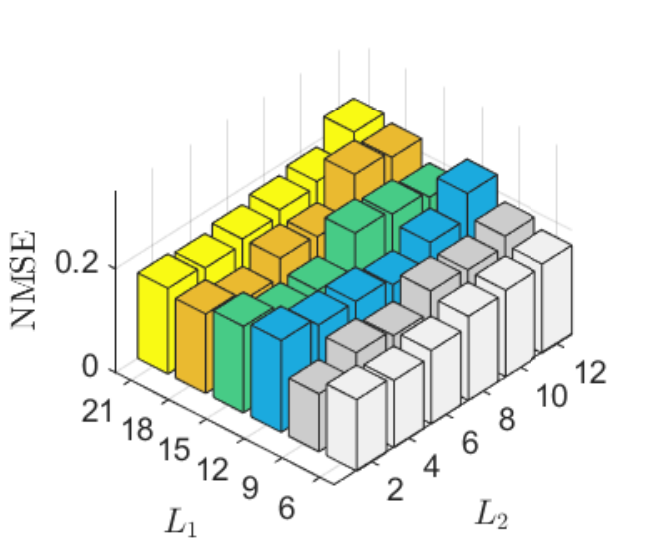}
		\label{20200625_1800}
	}
	\caption[fig2]{NMSE performance of JPSD estimator  ${\accentset{\circ}{\mathbf{\uptheta}}}_{{\bm x},\mathsf{GWM}}$ for: (a) Watts-Strogatz graph of $N = 100$ nodes and window pair $(K_1 = 7, K_2 = 5)$; (b) Watts-Strogatz graph of $N = 200$ nodes and window pair $(K_1 = 7, K_2 = 10)$; (c) Erdős–Rényi graph of $N = 100$ nodes and window pair $(K_1 = 7, K_2 = 5)$; and (d) Erdős–Rényi graph of $N = 200$ nodes and window pair $(K_1 = 7, K_2 = 10)$.}
	\label{fig:NMSE_vs_L1L2_SWG_N100}
\end{figure}

In our experiments, we divide EEG signals into frames of length $150$ ms as the realizations of JWSS processes. Then, estimated JPSD via~\eqref{GWMJPSD1837} with $\bm{A}_{\mathsf J} = \mathbf{I}_{NM}$ is used for the purpose of classification on EEG time-series on graph. It should be noted that our experiments using machine learning approach performed under the same experimental conditions. 
The common procedure between the two experiments are as follows. We first normalize the EEG dataset $\mathscr{D} = \{\bm{x}_k = \mathsf{vec}(\bm{X}_{k}): k \in {\llbracket 1,N_s \rrbracket}\}$ to the range $[0,1]$. 
After feature extraction, the feature matrix is normalized by exploiting the z-score technique which makes the values of each feature have zero-mean ($\mu_{\overline{\boldsymbol{z}}_{\ell}}=0$) and unit-variance ($\sigma_{\overline{\boldsymbol{z}}_{\ell}}^2=1$); the standardized feature matrix is denoted by $\overline{\boldsymbol{Z}} = [\overline{\boldsymbol{z}}_1, \ldots, \overline{\boldsymbol{z}}_{NM}]$ where
$
\overline{\boldsymbol{z}}_\ell = ({\boldsymbol{z}_\ell - {\mu_{\boldsymbol{z}_\ell}}\boldsymbol{1}})/{{\sigma_{\boldsymbol{z}_\ell}}},
$
and $\boldsymbol{z}_\ell$ is the $\ell$-th column of the original feature matrix ${\boldsymbol{Z}}$. Note that $\mu_{\boldsymbol{z}_\ell}$ and ${{\sigma_{\boldsymbol{z}_\ell}}}$ are the mean and variance of $\boldsymbol{z}_\ell$, respectively. 

\begin{figure}[t!]
	\centering
	\vspace*{-6mm}
	\hspace*{-1mm}
	\subfloat[]{
		\includegraphics[scale=0.68]{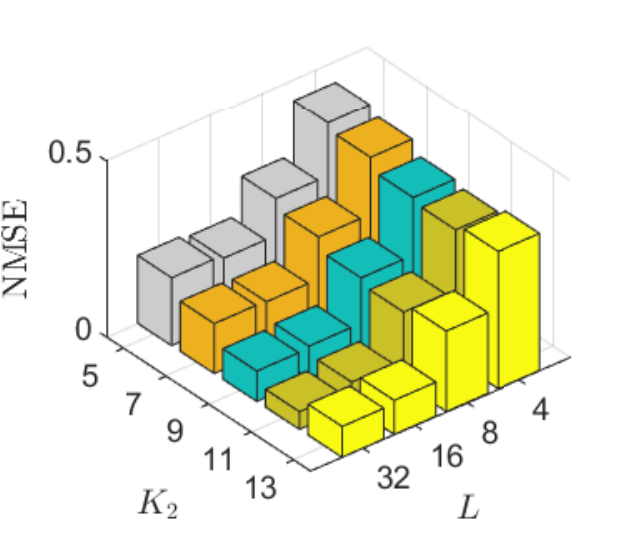}
		\label{20200217_0534}
	}
	\hspace*{-6mm}
	\subfloat[]{
		\includegraphics[scale=0.66]{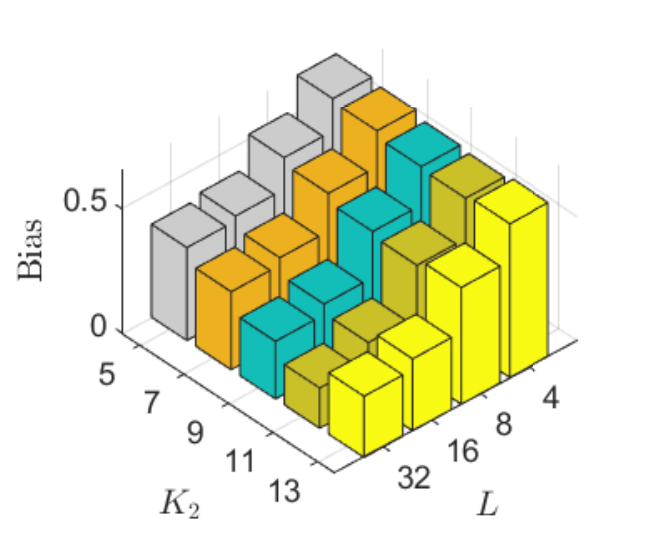}
		\label{20200217_0535}
	} 
	\hspace*{-7mm}
	\subfloat[]{
		\includegraphics[scale=0.69]{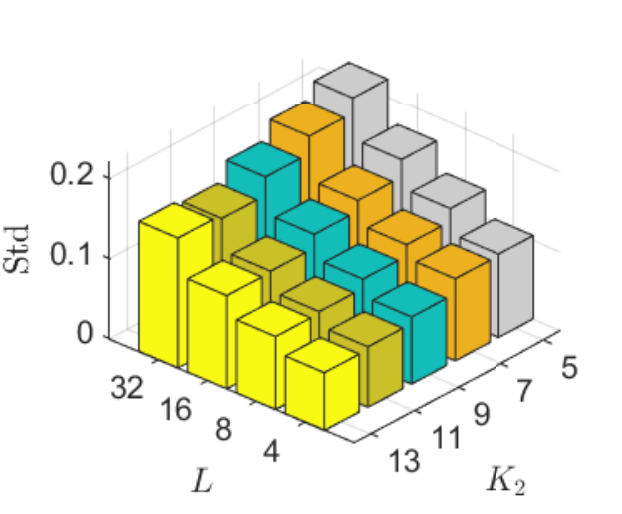}
		\label{20200217_0536}
	}
	\hspace*{-3mm}
	\subfloat[]{
		\includegraphics[scale=0.71]{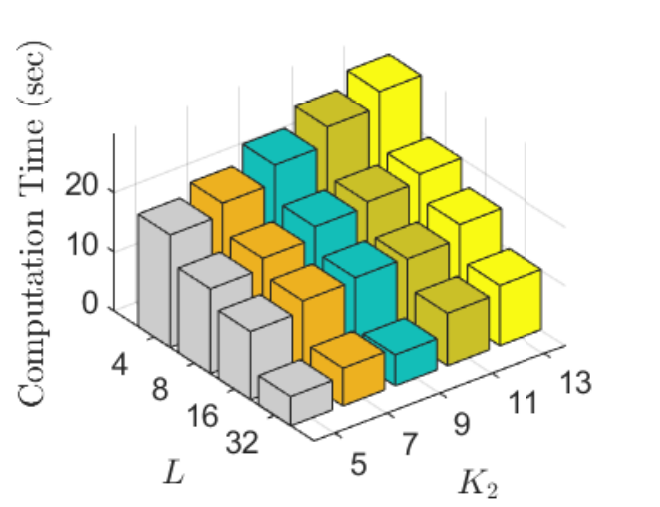}
		\label{20200217_0537}
	}
	\caption{Effect of the window length in discrete-time domain $L$ and number of graph windows $K_2$ on JPSD estimation using GWM: (a) NMSE; (b) Bias; (c) Std; and (d) Computation time, where a JFIR filter of degrees $L_1 = 7$, $L_2 = 4$ and a Erdős–Rényi graph with $N = 200$ vertices is used.}
	\label{fig:20200216_1252}
\end{figure}

\raggedright
\justify
{${\mathsf{\mathbf{ER~1\!:~Emotion~Recognition~From~EEG~Data.}}}$}
Recent attempts for emotion recognition using Electroencephalography (EEG) signals have demonstrated its effectiveness in human-machine interactions~\cite{Peixiang2019EEG}. In this section, we apply the concept of joint wide-sense stationarity for the challenging task of emotion recognition from brain EEG signals. The SEED-IV~\cite{Zheng2019EmotionMeter} is a publicly available EEG dataset obtained from $15$ subjects each participating in $3$ sessions, each session including $24$ trials. In each trial, every participant watched one out of $72$ movie clips while his/her EEG signals are collected via $62$-channel ESI NeuroScan System. The corresponding EEG channels are illustrated in Figure~\ref{20200921_1625} which is borrowed from~\cite{Zheng2019EmotionMeter}. The samples are categorized into four emotions as \textit{fear, happy, sad,} and \textit{neutral}. Our experiments are based on all the $N_s = 1080$ available samples in this dataset. As an initial preprocessing step, we filter out the frequencies less than $2$ Hz (also called \textit{slow-wave EEG activity or Delta wave}~\cite{greenfield2009reading}) and high frequency noise using a bandpass Butterworth filter ($2$--$150$~Hz). Moreover, a notch filter is applied to get rid of the $60$ Hz artifact which is a fundamental filtering step in EEG data analysis. Some studies have shown that the asymmetry in neuronal activities between the left and right hemispheres is useful for emotion recognition~\cite{dimond1976differing,Louis2001Frontal,Zhao2018Frontal}. Zhong~\textit{et al.}~\cite{Peixiang2019EEG} exploited this differential asymmetry information to initialize the adjacency matrix for developing the graph convolution network for emotion recognition. It is shown experimentally that the following set of channel pairs, denoted by ${\mathscr{E}}_{glb}$, balances the wiring cost and global efficiency, having particular importance in Emotion recognition~\cite{Peixiang2019EEG,Bullmore2012economy}: (FP1, FP2), (AF3, AF4), (F5, F6), (FC5, FC6), (C5, C6), (CP5, CP6), (P5, P6), (PO5, PO6), and (O1, O2). One may refer to~Figure~\ref{20200921_1625}. We build the brain graph based on the concept of \textit{local and global inter-channel relations} across all the EEG channels. Let ${\mathscr{E}}$ be set of all the edges connecting nodes in brain network. Then we define the weighted adjacency matrix $\mathbf{W}_{\mathsf G} = [w_{i,j}] \in \mathbb{R}_{+}^{62\times 62}$ based on the locations of EEG channels via a Gaussian kernel as follows:
\begin{align}\label{weightingMatrix}
w_{i,j} \coloneqq
\kappa \exp\Big(\frac{{-{\mathsf{dist}(i,j)}}}{2\gamma^2}\Big),
\end{align}
where $\gamma = 5.1$ is a scaling parameter, $\kappa = 2$ if $(i,j) \in {\mathscr{E}}_{glb}$ and $\kappa = 1$ if $(i,j) \in {\mathscr{E}} \setminus {\mathscr{E}}_{glb}$, and
\begin{align}\label{20200622_0745}
\mathsf{dist}(i,j) \coloneqq \|\bm{v}_i - \bm{v}_j\|_1
\end{align}
is the Manhattan distance between two EEG channels $i$ and $j$ with coordinate vectors $\bm{v}_i$ and $\bm{v}_j$, respectively. Note that the values for $\gamma$ and $\kappa$ are chosen empirically. In this modeling, we set $\kappa = 1$ for the local inter-channel relations, however, for the global connections we employ $\kappa = 2$ due to above-mentioned differential asymmetry information between right and left brain hemispheres in emotion recognition. 


\begin{figure}[t!]
	\vspace*{-0.4cm}
	\centering
	\subfloat[]{
	\includegraphics[scale=0.23]{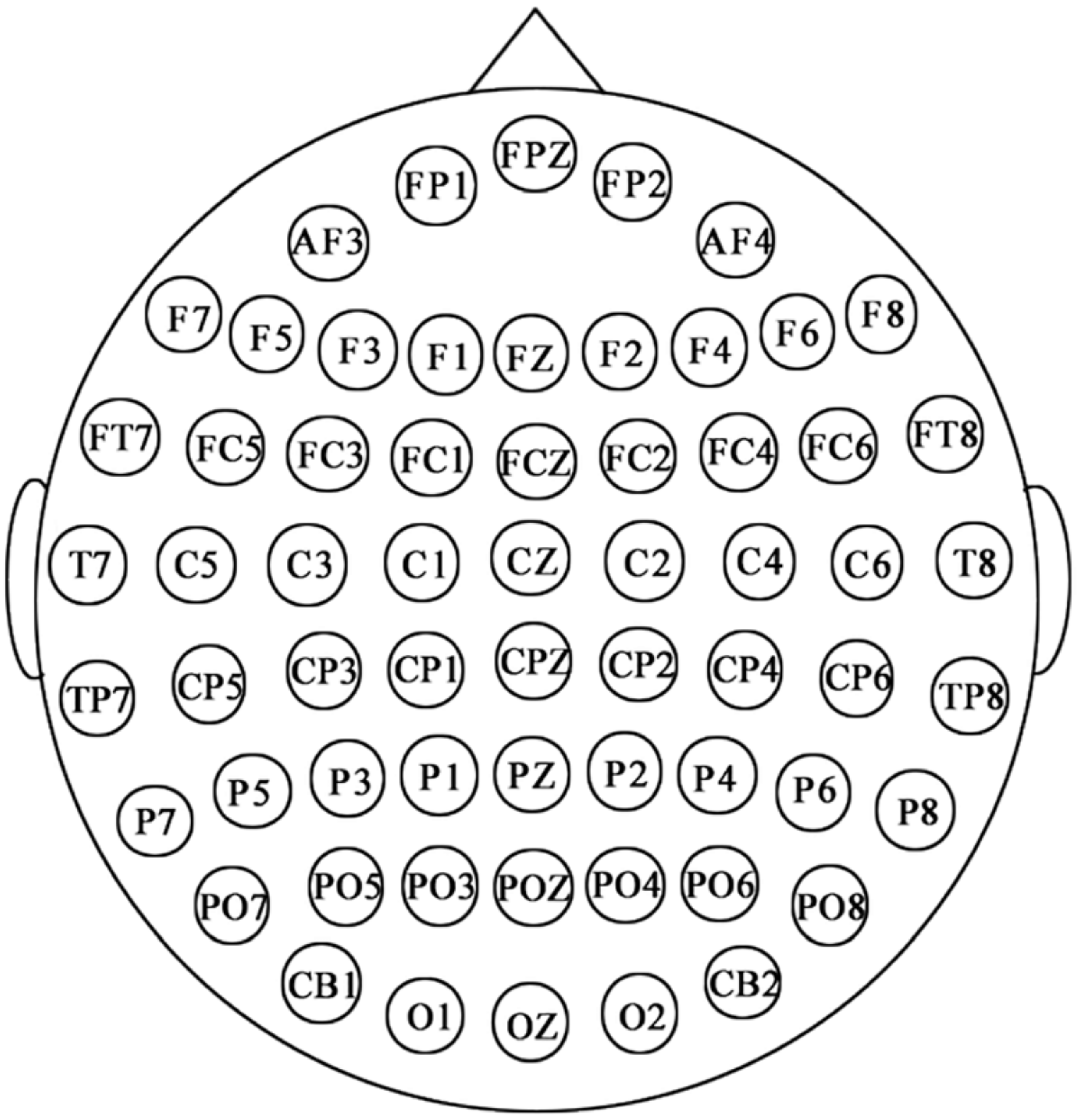}
	\label{20200921_1625}
	}
	\hspace*{10mm}
	\subfloat[]{
	\vspace*{-1cm}
	\includegraphics[scale=0.0275]{10_20EEG.pdf}
	\label{20200921_1626}
	} 
	\caption{The EEG layout of electrodes exploited in the collection of two different datasets: (a) $62$-channel ESI NeuroScan System; (b) International 10-20 system.}
	\label{fig:20200921_1627}
\end{figure}


Figure~\ref{fig:EEGEmotion} shows the considered Emotion recognition system from EEG signals using JPSD features.
Table~\ref{tab:tbl3} presents the correct classification rate (CCR) of emotion recognition obtained from three modelings of EEG signals namely TWSS, VWSS, and JWSS using the corresponding PSD estimates in each case as the feature sets. The methods have been tested on the SEED-IV dataset using SVM with radial basis kernel and stratified $10$-fold cross validation. Features obtained through TWSS, VWSS, and JWSS modeling reach up to $55.4\%$, $35.4\%$, and $59.5\%$ recognition accuracy (on average over all emotions), all using cross-validated PCA. Moreover, the obtained average of recognition accuracy over $30$ Monte~Carlo runs for the JWSS modeling is ${57.8}\pm {0.8}$ which is superior to the accuracy obtained from TWSS and VWSS modeling as $53.2 \pm 1.0$ and $32.4 \pm 0.9$, respectively\footnote{Note that due to the random split of data into training and test sets for the cross validation, performing Monte~Carlo simulation runs is necessary.}. Clearly, the feature set obtained from JWSS modeling (cf.~Definition~\ref{JWSSdef2} and~Theorem~\ref{20200622_1944}) performs better than VWSS and TWSS modeling in terms of recognition accuracy for emotion recognition from EEG signals. 

\begin{table}[t!]
	\caption{CCR of Emotion recognition from EEG signals for SEED-IV dataset using SVM} 
	\vspace*{0.1cm}
	\ra{1.3}
	\label{tab:tbl3}
	\centering
	\small 
	\begin{threeparttable}
		\begin{tabular}{c c *{4}{c} >{\columncolor{black!5}} c} 
			\toprule[1pt]
			& \multicolumn{4}{c}{CCR (\%)}  & \multicolumn{2}{c}{Accuracy (\%)}   \\ 
			\cmidrule{2-5} \cmidrule(l){6-7}
			{}   & Neutral  & Sadness & Fear & Happiness & Achievable  & Average \\
			\cmidrule{2-5} \cmidrule(l){6-7}  
			TWSS  & $63.7$ & $50.7$  & $54.4$ & $52.6$	& $55.4$ & {$53.2 \pm 1.0$} \\
			VWSS  & $27.4$ & $30.4$  & $38.2$ & $45.6$	& $35.4$ & {$32.4 \pm 0.9$} \\
			JWSS & $60.7$ & $54.1$  & $62.6$ & $60.7$    & ${\bm{59.5}}$ & {$\bm{57.4}\pm \bm{1.0}$} \\
			\bottomrule[1pt]
		\end{tabular}	
		
	\end{threeparttable} 
\end{table}



\raggedright
\justify
{${{\mathbf{ER~2\!:Alzheimer's~Disease~Recognition~From~EEG}}}$}
${\mathbf{Data.}}$
Alzheimer is a chronic neurodegenaritive disorder which appears in the form of progressive dementia with no specific cure. Due to its major public health challenge, analysis of EEG signals associated with AD is particularly important. Here, we aim at exploiting the PSD and JPSD as the features for AD recognition from EEG signals. For that, we use the AD dataset~\cite{Escudero2006Analysis} containing EEG signals taken from $12$ AD patients ($5$ men and $7$ women) and $11$ healthy control subjects with average age $72.8\pm 8.0$ recruited from the Alzheimer's Patients' Relatives Association of Valladolid. Note that there are $N_s = 663$ samples of EEG signals collected via the international $10$--$20$ system (cf. Figure~\ref{20200921_1626}). The $N=16$ channels are as follows: F3, F4, F7, F8, FP1, FP2, T3, T4, T5, T6, C3, C4, P3, P4, O1, and O2. To make this experiment more challenging and to show the effectiveness of JPSD in noisy data, we add additive white Gaussian noise to the raw data and do the classification afterwards. We define the weighted adjacency matrix $\mathbf{W}_{\mathsf G} = [w_{i,j}] \in \mathbb{R}_{+}^{16\times 16}$ based upon the locations of EEG channels via using~\eqref{weightingMatrix} with $\gamma = 6$ and $\kappa = 1$.

The methods have been tested on the AD dataset by SVM classifier with radial basis kernel and stratified $10$-fold cross validation. The system for AD recognition is the same as emotion recognition system which is shown in Figure~\ref{fig:EEGEmotion}. Figure~\ref{fig:CCRSNR} depicts the CCR of AD recognition for a range of signal to noise ratio (SNR) associated with features obtained from TWSS, VWSS, and JWSS modeling. Here, the CCR is obtained by averaging recognition accuracy over $50$ Monte~Carlo simulation runs and the corresponding Std in each case is shown by the bounded vertical bar. Clearly, JWSS modeling has superior performance to the TWSS and VWSS for all the tested SNRs and demonstrates higher robustness of JPSD features against noise. It is worth noting that impact of temporal information for AD recognition is much more significant than the spatial information. Also note that as the SNR increases, the Std of recognition accuracy also decreases.

\section{Additional Declarations}

\raggedright
\justify
{\bf Availability of Data.}
The data that we used in our experiments are as follows:
\begin{itemize}
	\item SEED~IV~\cite{Zheng2019EmotionMeter}: The dataset concerning emotion recognition from EEG signals is available at: {\bf\small{\texttt{\url{http://bcmi.sjtu.edu.cn/~seed/seed-iv.html}}}}. 
	\item AD dataset~\cite{Escudero2006Analysis}: The dataset concerning AD recognition from EEG signals is available at: {\bf\small{\texttt{\url{https://osf.io/jbysn/}}}}. 
\end{itemize}

\begin{note}
	We need to specify that the datasets concerning with neurological disorders are in general not publicly available, however, this framework can potentially be used in the case of availability of challenging datasets concerning with patients in different stages of such diseases. 
\end{note}

\raggedright
\justify
{\bf Code.}
The source code of our simulations and experiments will be publicly available at:
{\bf\small{\texttt{\url{https://github.com/amin-jalili/mfdssp/projects/1}}}}.

\begin{figure*}[t!]
	\vspace*{-0.2cm}
	\centering
	\includegraphics[width=1\linewidth]{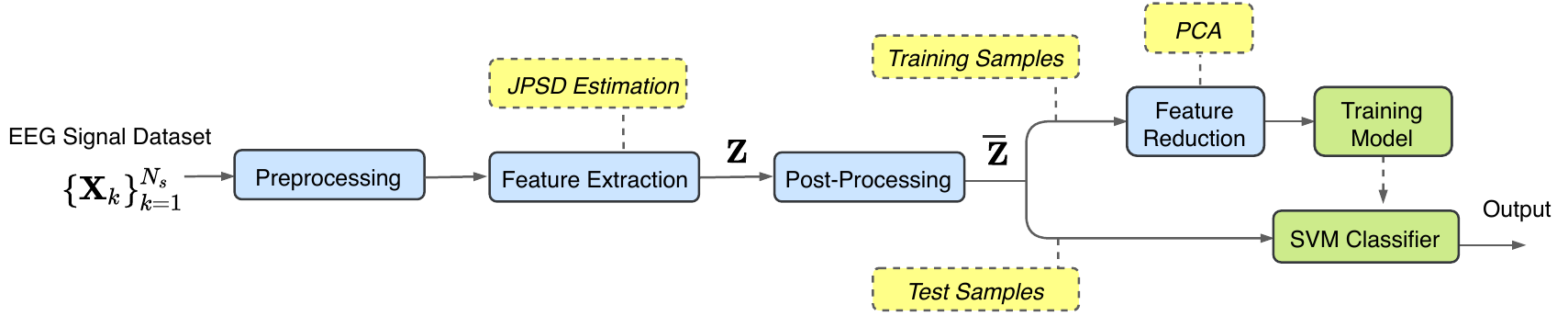}  
	\caption{Emotion and AD recognition system from EEG signals using the JPSD features}
	\label{fig:EEGEmotion} 
\end{figure*}

%

\raggedright
\justify
{\bf Abbreviations.}
DFT: \textit{Discrete-Time Fourier Transform}; 
IDFT: \textit{Inverese DFT}; 
GFT: \textit{Graph Fourier Transform}; 
IGFT: \textit{Inverse GFT}; 
JFT: \textit{Joint (Time-Vertex) Fourier Transform};
IJFT: \textit{Inverse JFT};
GTO: \textit{Graph Transition Operator}; 
JTO: \textit{Joint Transition Operator}; 
JWSS: \textit{Joint Wide-Sense Stationary}; 
VWSS: \textit{Vertex Wide-Sense Stationary}; 
MVWSS: \textit{Multivariate VWSS};
TWSS: \textit{Temporal Wide-Sense Stationary}; 
MTWSS: \textit{Multivariate TWSS}; 
PSD: \textit{Power Spectral Density}; 
JPSD: \textit{Joint Power Spectral Density}; 
GBM: \textit{Generalized Bartlett Method};
GWM: \textit{Generalized Welch Method};
AD: \textit{Alzheimer's disease}.

\section{Conclusion}

We have presented a generalized framework for modeling the stochastic time-varying graph signal as a JWSS process via the proposed bivariate isometric JTO. It was shown that the resulting notion of wide-sense stationarity via bivariate JTO in the time-vertex domain is more general than defining stationarity via isometric GTO on joint graph. This approach can be applied to the JPSD estimation of time-series on graph, from which one can use the resulting JPSD as the features for machine learning based applications. Then we presented the generalized Welch estimator for JPSD estimation, supported by some simulation results. Eventually, we tested the applicability of JPSD features for the classification of time-series on graph including Emotion and AD recognition from EEG signals by modeling them as JWSS processes. The experimental results demonstrated that the proposed JWSS modeling with JPSD features outperforms the classical TWSS and VWSS modelings with PSD and GPSD as features, respectively. In spite of a simple graph learning method used for the classification of EEG data, more effective methods will be our future study for upgrading emotion recognition in the GSP framework. Moreover, as another future study, it is interesting to exploit the JPSD features to classify more challenging datasets with multiple classes associated with different stages of neurological disorders like AD, Parkinson, and Epilepsy.  Finally, the JPSD features can potentially be used for other real world applications associated with the classification of time-series on graph, where the JWSS is a suitable modeling.

\appendix
\section{Proof of Remark~\ref{20200614_1227}}
\label{AppendixI}
Let $\mathcal{D}_{\mathsf C} \coloneqq \partial_t$ denote the differentiation w.r.t. $t$ and $\widehat{x}(\xi)$ be the Fourier transform of $x(t)$. Using the Taylor series expansion 
\begin{align}\label{shiftCSP1}
(\mathcal{T}_{\mathsf C}^{\tau} x)(t) = x(t-\tau) = \sum_{k=0}^{\infty} \frac{1}{k!} \left(-\tau \mathcal{D}_{\mathsf C} \right)^k x(t). 
\end{align}	
From the theory of Fourier transform, it is easy to verify the following property as
$
\mathcal{F}_{\mathsf C} \mathcal{D}_{\mathsf C} x(t) =  \mi2\pi \xi \widehat{x}(\xi)
$~\cite{Oppenheim1989Discrete}.
In a compact notation one can write
$
\mathcal{F}_{\mathsf C} \mathcal{D}_{\mathsf C} = \mi \mathcal{M}_{\mathsf C}\mathcal{F}_{\mathsf C}
$
where 
\begin{align}\label{MultplierOp1940}
\mathcal{M}_{\mathsf C}\widehat{x}(\xi) \coloneqq 2\pi \xi\widehat{x}(\xi),
\end{align}
is an explicit multiplier. Then one can obtain 
\begin{align}\label{diferentiationOp}
\mathcal{D}_{\mathsf C} = \mi \mathcal{F}_{\mathsf C}^{-1}\mathcal{M}_{\mathsf C}\mathcal{F}_{\mathsf C}.
\end{align}
Equation~\eqref{shiftCSP1} together with~\eqref{diferentiationOp} implies that
\begin{align}\label{shiftCSP2}
\begin{split}
\mathcal{T}_{\mathsf C}^{\tau}  &= 1- \frac{\tau}{1!} \mathcal{D}_{\mathsf C} + \frac{\tau^{2}}{2!} \mathcal{D}_{\mathsf C}^{2} - \cdots \\
&= \mathcal{F}_{\mathsf C}^{-1} \big(1- \mi \frac{\tau}{1!}(2\pi\xi) +\mi^{2} \frac{\tau^{2}}{2!} (2\pi\xi)^{2} - \cdots \big) \mathcal{F}_{\mathsf C},
\end{split}
\end{align}	
which clearly reduces to~\eqref{shiftOpContiDomain}.
\hfill $\blacksquare$

\begin{figure}[t!]
	\centering
	\vspace*{-2mm}
	\hspace*{-4mm}
	\includegraphics[scale=0.73]{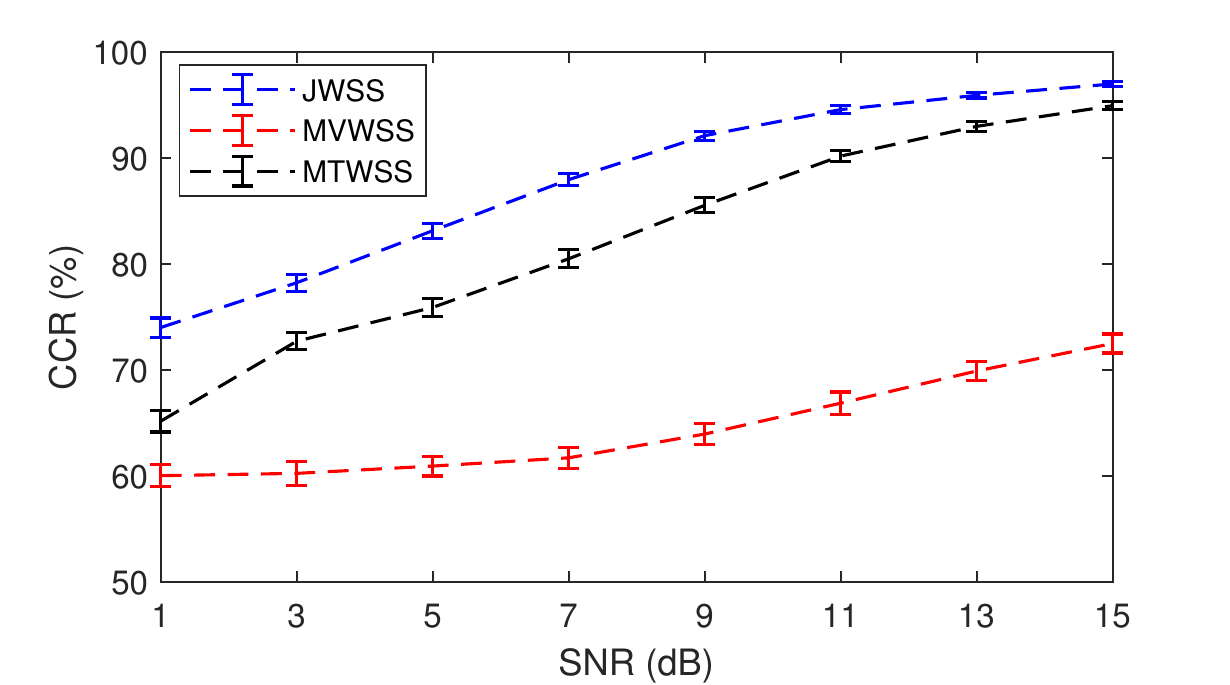}
	\caption[fig1]{CCR performance versus SNR of AD recognition from EEG signals using TWSS, VWSS, and JWSS modeling where SVM with $10$-fold cross validation is used.}
	\label{fig:CCRSNR}
\end{figure}




\begin{thebibliography}{10}
	\expandafter\ifx\csname url\endcsname\relax
	\def\url#1{\texttt{#1}}\fi
	\expandafter\ifx\csname urlprefix\endcsname\relax\def\urlprefix{URL }\fi
	\expandafter\ifx\csname href\endcsname\relax
	\def\href#1#2{#2} \def\path#1{#1}\fi
	
	\bibitem{Amin2020stationarity}
	A.~{Jalili}, C.-Y. {Chi}, {Stationarity of Time-Series on Graph via Bivariate
		Translation Invariance}, arXiv e-prints:2004.00298 (April 2020).
	
	\bibitem{Shuman2013Emerging}
	{D. I Shuman}, S.~K. Narang, P.~Frossard, A.~Ortega, P.~Vandergheynst, The
	emerging field of signal processing on graphs: {E}xtending high-dimensional
	data analysis to networks and other irregular domains, IEEE Signal Process.
	Mag. 30~(3) (2013) 83--98.
	
	\bibitem{Chung1997Spectral}
	F.~R.~K. Chung, {Spectral Graph Theory}, Vol.~92 of CBMS Regional Conference
	Series in Mathematics, American Mathematical Society, 1997.
	
	\bibitem{HAMMOND2011129Wavelets}
	D.~K. Hammond, P.~Vandergheynst, R.~Gribonval, Wavelets on graphs via spectral
	graph theory, Applied and Computational Harmonic Analysis 30~(2) (2011)
	129--150.
	
	\bibitem{Sandryhaila2013Discrete}
	A.~Sandryhaila, J.~M.~F. Moura, Discrete signal processing on graphs, IEEE
	Trans. Signal Processing 61~(7) (2013) 1644--1656.
	
	\bibitem{Puschel2008Algebraic}
	M.~{Püschel}, J.~M.~F. {Moura}, Algebraic signal processing theory: Foundation
	and 1-d time, IEEE Trans. Signal Processing 56~(8) (2008) 3572--3585.
	
	\bibitem{STANKOVIC2020Vertex}
	L.~Stanković, D.~Mandic, M.~Daković, B.~Scalzo, M.~Brajović, E.~Sejdić,
	A.~G. Constantinides, Vertex-frequency graph signal processing: A
	comprehensive review, Digital Signal Processing 107 (2020) 102802.
	
	\bibitem{Shuman2016Vertex}
	D.~I. Shuman, B.~Ricaud, P.~Vandergheynst, Vertex-frequency analysis on graphs,
	Applied and Computational Harmonic Analysis 40~(2) (2016) 260--291.
	
	\bibitem{perraudin2017graph}
	N.~Perraudin, Graph-based structures in data science: Fundamental limits and
	applications to machine learning, Ph.D. thesis, Ecole Polytechnique Federale
	de Lausanne, Lausanne, Switzerland (2018).
	
	\bibitem{Bronstein2017Geometric}
	M.~M. {Bronstein}, J.~{Bruna}, Y.~{LeCun}, A.~{Szlam}, P.~{Vandergheynst},
	Geometric deep learning: Going beyond {E}uclidean data, IEEE Signal
	Processing Magazine 34~(4) (2017) 18--42.
	
	\bibitem{perraudin2018Global}
	N.~Perraudin, B.~Ricaud, D.~I. Shuman, P.~Vandergheynst, Global and local
	uncertainty principles for signals on graphs, APSIPA Trans. Signal and
	Information Processing 7, e3 (2018).
	
	\bibitem{Giannakis2018Topology}
	G.~B. {Giannakis}, Y.~{Shen}, G.~V. {Karanikolas}, Topology identification and
	learning over graphs: Accounting for nonlinearities and dynamics, Proceedings
	of the IEEE 106~(5) (2018) 787--807.
	
	\bibitem{Dong2019Learning}
	X.~{Dong}, D.~{Thanou}, M.~{Rabbat}, P.~{Frossard}, Learning graphs from data:
	A signal representation perspective, IEEE Signal Processing Magazine 36~(3)
	(2019) 44--63.
	
	\bibitem{Dong2014Multi}
	X.~Dong, Multi-view signal processing and learning on graphs, Ph.D. thesis,
	Ecole Polytechnique Federale de Lausanne, Lausanne, Switzerland (2014).
	
	\bibitem{Grassi2018Time}
	F.~{Grassi}, A.~{Loukas}, N.~{Perraudin}, B.~{Ricaud}, A time-vertex signal
	processing framework: Scalable processing and meaningful representations for
	time-series on graphs, IEEE Trans. Signal Processing 66~(3) (2018) 817--829.
	
	\bibitem{Girault2015Stationary}
	B.~Girault, Stationary graph signals using an isometric graph translation, in:
	23rd European Signal Processing Conference (EUSIPCO), 2015, pp. 1516--1520.
	
	\bibitem{Girault2015Trans}
	B.~Girault, P.~Gonçalves, {\'E}.~Fleury, Translation on graphs: An isometric
	shift operator, IEEE Signal Processing Letters 22~(12) (2015) 2416--2420.
	
	\bibitem{Perraudin2017Stationary}
	N.~Perraudin, P.~Vandergheynst, Stationary signal processing on graphs, IEEE
	Trans. Signal Processing 65~(13) (2017) 3462--3477.
	
	\bibitem{Marques2017Stationary}
	A.~G. Marques, S.~Segarra, G.~Leus, A.~Ribeiro, Stationary graph processes and
	spectral estimation, IEEE Trans. Signal Processing 65~(22) (2017) 5911--5926.
	
	\bibitem{Loukas2017JStationary}
	A.~Loukas, N.~Perraudin, Stationary time-vertex signal processing, EURASIP
	Journal on Advances in Signal Processing (Nov. 2019).
	
	\bibitem{SEGARRA2018Statistical}
	S.~{Segarra}, S.~P. {Chepuri}, A.~G. {Marques}, G.~{Leus}, Chapter 12 -
	{S}tatistical {G}raph {S}ignal {P}rocessing: Stationarity and spectral
	estimation, in: P.~M. Djurić, C.~Richard (Eds.), Cooperative and Graph
	Signal Processing, Academic Press, 2018, pp. 325 -- 347.
	
	\bibitem{Perraudin2017Towards}
	N.~Perraudin, A.~Loukas, F.~Grassi, P.~Vandergheynst, Towards stationary
	time-vertex signal processing, in: 2017 IEEE International Conference on
	Acoustics, Speech and Signal Processing (ICASSP), 2017, pp. 3914--3918.
	
	\bibitem{Isufi2019Forecasting}
	E.~{Isufi}, A.~{Loukas}, N.~{Perraudin}, G.~{Leus}, Forecasting time series
	with {VARMA} recursions on graphs, IEEE Trans. Signal Processing 67~(18)
	(2019) 4870--4885.
	
	\bibitem{Higham2008Functions}
	N.~J. Higham, Functions of Matrices: {Theory} and Computation, Society for
	Industrial and Applied Mathematics, Philadelphia, PA, USA, 2008.
	
	\bibitem{vetterli2014Foundations}
	M.~Vetterli, J.~Kovačević, V.~K. Goyal, Foundations of Signal Processing,
	Cambridge University Press, 2014.
	
	\bibitem{Loukas2016Frequency}
	A.~{Loukas}, D.~{Foucard}, Frequency analysis of time-varying graph signals,
	in: Proc. 2016 IEEE Global Conference on Signal and Information Processing
	(GlobalSIP), 2016, pp. 346--350.
	
	\bibitem{Sandryhaila2014Discrete}
	A.~{Sandryhaila}, J.~M.~F. {Moura}, Discrete signal processing on graphs:
	Frequency analysis, IEEE Trans. Signal Processing 62~(12) (2014) 3042--3054.
	
	\bibitem{Bach2004Learning}
	F.~R. {Bach}, M.~I. {Jordan}, Learning graphical models for stationary time
	series, IEEE Trans. Signal Processing 52~(8) (2004) 2189--2199.
	
	\bibitem{girault2015Signal}
	B.~Girault, Signal processing on graphs - contributions to an emerging field,
	Ph.D. thesis, {Ecole normale sup{\'e}rieure de lyon - ENS LYON}, France
	(2015).
	
	\bibitem{Horn1985Matrix}
	R.~A. Horn, C.~R. Johnson (Eds.), Matrix Analysis, Cambridge University Press,
	USA, 1985.
	
	\bibitem{bollobas2001Random}
	B.~Bollobás, Random Graphs, 2nd Edition, Cambridge Studies in Advanced
	Mathematics, Cambridge University Press, 2001.
	
	\bibitem{Watts1998Collective}
	D.~J. Watts, S.~H. Strogatz, {Collective dynamics of small-world networks},
	Nature 393~(6684) (1998) 440--442.
	
	\bibitem{Peixiang2019EEG}
	P.~Zhong, D.~Wang, C.~Miao, {EEG}-based emotion recognition using regularized
	graph neural networks, arXiv e-prints:1907.07835 abs/1907.07835 (Jul. 2019).
	
	\bibitem{Zheng2019EmotionMeter}
	W.~{Zheng}, W.~{Liu}, Y.~{Lu}, B.~{Lu}, A.~{Cichocki}, Emotionmeter: A
	multimodal framework for recognizing human emotions, IEEE Trans. Cybernetics
	49~(3) (2019) 1110--1122.
	
	\bibitem{greenfield2009reading}
	L.~Greenfield, J.~Geyer, P.~Carney, Reading EEGs: A Practical Approach, LWW
	medical book collection, Lippincott Williams \& Wilkins, 2009.
	
	\bibitem{dimond1976differing}
	S.~J. Dimond, L.~Farrington, P.~Johnson, Differing emotional response from
	right and left hemispheres, Nature 261~(5562) (1976) 690--692.
	
	\bibitem{Louis2001Frontal}
	L.~A. Schmidt, L.~J. Trainor, Frontal brain electrical activity ({EEG})
	distinguishes valence and intensity of musical emotions, Cognition and
	Emotion 15~(4) (2001) 487--500.
	
	\bibitem{Zhao2018Frontal}
	G.~Zhao, Y.~Zhang, Y.~Ge, Frontal {EEG} asymmetry and middle line power
	difference in discrete emotions, Frontiers in Behavioral Neuroscience 12
	(2018) 225.
	
	\bibitem{Bullmore2012economy}
	E.~Bullmore, O.~Sporns, The economy of brain network organization, Nature
	reviews. Neuroscience 13 (2012) 336--49.
	
	\bibitem{Escudero2006Analysis}
	J.~Escudero, D.~Ab{\'a}solo, R.~Hornero, P.~Espino, M.~L{\'o}pez, Analysis of
	{E}lectroencephalograms in {A}lzheimer's disease patients with multiscale
	entropy, Physiological Measurement 27~(11) (2006) 1091--1106.
	
	\bibitem{Oppenheim1989Discrete}
	A.~V. Oppenheim, R.~W. Schafer, Discrete-time Signal Processing, Prentice Hall,
	1989.
	
\end{thebibliography}

\end{document}